\documentclass[conference]{IEEEtran}
\usepackage{cite}
\usepackage[pdftex]{graphicx}
\usepackage[cmex10]{amsmath}
\usepackage{amssymb}

\usepackage{times}
\usepackage{array}
\usepackage[tight,footnotesize]{subfigure}
\usepackage[font=footnotesize]{subfig}
\usepackage{graphicx}
\usepackage{mathtools}
\usepackage{microtype}
\usepackage{amsmath}

\usepackage{alltt}
\usepackage{shadethm}
\newshadetheorem{comment}{Comment}

\def\boxx{{\vcenter{\vbox{\hrule height.3pt
          \hbox{\vrule width.3pt height6pt
          \kern6pt\vrule width.3pt}\hrule height.3pt}}\;}}

\def\impos{{\;\vcenter{\hbox{\rule{5mm}{0.2mm}}} \vcenter{\hbox{\rule{1.5mm}{1.5mm}}} \;}}

\def\lrarrow{\leftrightarrow \kern-8pt \rightarrow}

\def\2{\frac{1}{2}}

\def\beq{\begin{eqnarray}}
\def\eeq{\end{eqnarray}}
\def\2{\frac{1}{2}}

\newtheorem{lemma}{Lemma}
\newtheorem{principle}{Principle}
\newtheorem{proposition}{Proposition}
\newtheorem{definition}{Definition}
\newtheorem{remark}{Remark}

\def\lrarrow{\leftrightarrow \kern-8pt \rightarrow}

\def\frightarrow{\rightarrow \kern-11pt /~~}
\def\reducesto{\simeq \kern -3pt >}
\def\intersection{\cap}

\begin{document}
\newcommand{\strust}[1]{\stackrel{\tau:#1}{\longrightarrow}}
\newcommand{\trust}[1]{\stackrel{#1}{{\rm\bf ~Trusts~}}}
\newcommand{\promise}[1]{\xrightarrow{~#1~}}
\newcommand{\revpromise}[1]{\xleftarrow{#1} }
\newcommand{\assoc}[1]{{\xrightharpoondown{#1}} }
\newcommand{\rassoc}[1]{{\xleftharpoondown{#1}} }
\newcommand{\imposition}[1]{\stackrel{~#1~}{\impos}}
\newcommand{\scopepromise}[2]{\xrightarrow[#2]{#1}}
\newcommand{\handshake}[1]{\xleftrightarrow{#1} \kern-8pt \xrightarrow{} }
\newcommand{\cpromise}[1]{\stackrel{#1}{\frightarrow}}
\newcommand{\policy}{\stackrel{P}{\equiv}}
\newcommand{\field}[1]{\mathbf{#1}}
\newcommand{\bundle}[1]{\stackrel{#1}{\Longrightarrow}}

\title{Continuous Integration of Data Histories\\into Consistent Namespaces}

\author{Mark Burgess\\ChiTek-i AS, Norway\\~\\and\\~\\Andr\'as Gerlits\\Dianemo Ltd, Ireland\\~\\\today}
\maketitle
\IEEEpeerreviewmaketitle

\renewcommand{\arraystretch}{1.2}

\begin{abstract}
  We describe a policy-based approach to the scaling of shared data
  services, using a hierarchy of calibrated data pipelines to automate
  the continuous integration of data flows.  While there is no unique
  solution to the problem of time order, we show how to use a fair
  interleaving to reproduce reliable `latest version' semantics in a
  controlled way, by trading locality for temporal resolution.  We
  thus establish an invariant global ordering from a spanning tree
  over all shards, with controlled scalability.  This forms a
  versioned coordinate system (or versioned namespace) with consistent
  semantics and self-protecting rate-limited versioning, analogous to
  publish-subscribe addressing schemes for Content Delivery Network
  (CDN) or Name Data Networking (NDN) schemes.
\end{abstract}

\hyphenation{before}
\hyphenation{immun-ology}


\tableofcontents

\section{Introduction} 

The scaling and consistency of distributed information systems are two
sides of an ongoing narrative in Computer Science about consistent and
reliable access to data\cite{consensus4,andras,andraspatent1}.  The two subjects go hand in hand,
principally because increased scale means increased distribution over
space and time, and thus greater uncertainty about when events
occurred. Serial multiplexing is time-sharing, and shared memory
policy is memory or storage space-sharing.  Movement of data,
physically and virtually, connects change over different forms of
space and time through {\em process} trajectories\cite{virtualmotion}.
Where such trajectories come together, they must be interleaved by
defining a policy for their continuous integration\cite{koalja}.

Increased scale implies increased exposure to non-deterministic
environmental influences, including the effects of faults, latencies, and contention
for shared memory (races).  Processes, which attempt to assimilate and
calibrate data from multiple sources into a singular consistent view
for all, struggle against the limitations of communication and scale.
The semantics of this assimilation are typically left undefined in literature,
as if obvious and universal, but aggregation is a form of data pipelining
and there are several alternatives. This is where we shall try to do better.

A number of familiar tools are in widespread use, for defining and
ensuring data consistency, and after fifty years there's an
understandably large literature on the topic, see for example
\cite{paxos,raft,zookeeper,crdt,consensus1,consensus2,consensus3,consensus4}.
A common philosophy is to try to conceal the side-effects of scale
behind infrastructure, which simulates a single computer by brute
force, but this is both expensive and increasingly problematic in the
age of carbon budgets and energy awareness\cite{spanner}.  Often these
involve a high cost in terms of communication and memory; however,
they do not scale favourably to either the very large or the very
small.

A key part of the difficulty is designing a scalable data recorder
lies in determining the relative order of events that happen within
independent processes, sourced from different locations and
integrating them into a common history (scaling causality itself). Events without causal
dependencies have no natural order, and---while we might try to define
one by means of a global clock---a clock only has meaning within a
limited scope, where the time it takes to read the clock is
insignificant compared to the rate of change of the processes
themselves. At current process speeds, that assumption fails quickly
in a globalized world.

Another part of the difficulty in scaling has to do with the
insistence on `push-based' update models for the imposition of data
transactions. It's hard to overstate how deeply ingrained this
thinking is in IT. Push methods rely on purely reactive methods, which
are designed on the implicit assumption of a sender update process
that's sparse compare to the available schedule of the receiver.  For
saturated update streams, the data pipelines favour a scheduled `pull'
processing model, which optimizes the scarce time resource at the receiving
end, and avoids the unnecessary contention of `push'
semantics\cite{treatise2}\footnote{The distinction is analogous to that between Ethernet
and ATM or MPLS, in networking, for instance.}.  The interplay between the dynamic and
semantic issues then has to be balanced: how should we dimension and
interpret the intended behaviour of a system?  These are not questions
that can be answered once and for all cases, so there is a tendency to
succumb to ad hoc approaches.  Such ad hoc solutions lead to
uncertainties and reliability issues.

The challenge faced by any data service is to present a consistent
view of updates, and historical timelines to all users, regardless of
where they might be located.  This issue can only be solved by
curating a solution, since there is no absolute notion of time that
applies to all agents within a distributed system. Thus, several
different solutions may be possible, and different solutions may be
more suitable for some clients than for others.

By now the infamous `CAP conjecture'\footnote{The conjecture was
  misrepresented as the CAP Theorem, though no proper theorem has ever
  been expressed to cover the claims. The statement of the claim has
  gradually changed over time to try to transform it into a provable
  theorem. However, it's value lies in the rough observation of a
  truism rather than in a provable statement.} dominates many popular
discussions about such consistency of access to data. The CAP is the
poster-child for how ambiguity arises due to incomplete
semantics\cite{cap}. In practice, it's impossible to give meaning to a
single unique view of data for all clients.  All such attempts at
consistency involve some kind of compromise to manipulate space and
time, either by imposing a wait, a limit on observability, or by
accepting variance in data arrival times to different client. One
variance that's frowned upon is differences in the perceived order of
events, and we take it as an axiom that preserving the intended order
of data is a basic promise that databases must keep.  We resolve this
issue by defining an adaptive scale-dependent tradeoff---a kind of
data prism---in which data flows are `lensed' into a focused beam of
updates, by fair weighted interleaving.

Lamport was early in recognizing the issue of process relativity, by
defining a notion of consistency for distributed
information\cite{vectorclocks}. He used this to build Paxos, a widely
adopted quorum solver---for determining outcomes according to a fixed
policy\cite{paxos}. It builds on Gray et al's two-phase commit
protocol\cite{twophasecommit}.  Since agreement is the basis of
semantics in any collection of processes, it's natural that this
problem would dominate concerns about distributed intent.  However,
the quorum problem is about semantics of agreement, and is distinct
from the purely dynamical problem of observing data consistently.  The
vector clock approach Fridge\cite{fridge} identifies similar issues.
The ability to be certain of outcomes by asynchronous messaging is
subject to the so-called FLP no-go theorem \cite{flp}. The notion of
rigid entanglement allows us to side step this by maintaining
synchronous communication\cite{paulentangle} for part of the process.

\section{Overview}

In this note, we describe a generic spacetime approach to resolving
the order and causal consistency of data, applied to storage,
processing, replication, and retrieval\cite{andras,andraspatent1}. Unlike quorum solvers, which
define correctness by vote, our definition of correct version is based
on deterministic version numbering, and causal interleaving policy.
Data remain in place at their point of collection and a hierarchy of
associative `index' pointers is generated in real time to assign
coordinates to consistent slices of data. These become
observable globally after a scale-dependent
delay.  We describe the dynamical process by which addressable
namespace coordinates are assigned to changes, in order to integrate
data both dynamically and semantically into past and future `cones'.
This is a shared view, for all clients, based on a fair interleaving
of random spacetime arrivals (see figure \ref{cones}).

An important side effect of this approach is the impact it has on
cache replication. Replication can be decoupled from other processes
to maximize availability and the minimize the impact of partitions.
The replication of multi-version state can be acquired from an
authoritative {\em source}, by means of a change log. This is a
standard approach used by journaling filesystems and for rewinding
database snapshots. The more difficult problem occurs when we want to
replicate state from multiple sources, and at different locations, in
order to a reconstruct a consistent remote cache. This is where we
apply the concept of a distributed clock to construct a `Just In Time'
partial cache image that preserves the semantics of the clock. The
proposed clock therefore preserves a notion of the original order,
i.e. that which was `intended' by the original sources, from as local
a region as the required data allow.

Our proposed solution is based on `proper time' counting (the root
concept behind behind tensor clocks\cite{matrixclock,intervalclock}),
and can be scaled from very small to very large assemblies of
collaborating agents. The dynamical trade-offs can be controlled
deterministically by exploiting a hierarchical semantic coordinate
system\cite{spacetime1,andras}, and so-called `once only' semantics of
updates can be handled (in the usual manner) by designing for
convergent (idempotent) name coordinates\cite{burgessC11}.  The
automatic assignment of versions to key values is thus basically in
the spirit of `Continuous Integration' (well known from Software
Engineering), and is somewhat similar in spirit to addressability for
Name Data Networking \cite{ndn0,ndn} or Content Delivery Network
publisher-subscriber schemes.

The plan for the paper is as follows.  We begin by looking at a
causal collision-resolution mechanism for overlapping transactions involving a
single key-value. This acts as a reference and point of departure.  Then we
proceed to scale this process, in spacetime, by decoupling the
processes within the entire data pipeline: from change capture at the
edge of the network (writes), to publication of change (commits),
subscription to data channels (versioned reads), and finally
replication and caching on demand.

The observability of committed changes as publish-subscribe channels
(sometimes called `liveness' of updates), is integrated into a shared
absolute coordinate system, by a process of statistical interleaving,
which defines a consistent and intuitive view of the past. We infer
that this may be defined to be as close to what was {\em intended} by
the original writig processes as possible.  Finally, we comment on the
tradeoffs of our approach, specifically the maximum rate at which
data can become observable by the whole hierarchy and the implications
of this for distributed computations.

Our analysis in this note is based on Promise
Theory\cite{promisebook,treatise2} and its derived model of Semantic
Spacetime\cite{spacetime1} (see the appendix for a summary of
definitions and notations).  The components of our model are all known
(at least in principle) in the literature, and we hope to clarify
their composition into a distributed collaboration to distill a
generic solution.

\section{Data configuration space and background}

Data storage management is essentially a form of rapid incremental
configuration management on demand. The scope of data management is
usually much greater than one would normally expect for
`configuration' changes (a term usually associated with more slowly
varying permanent infrastructure). By now, the accepted approach for
configuration is to ensure an invariant state through monotonic and
idempotent state convergence, as proven by Burgess
in\cite{burgessC1,burgessC11}, building on Shannon's error correction
theorem\cite{shannon1}.  Convergent correctness in a data store may be
viewed as a version of error correction over a single policy domain.
Meanwhile, contention-free handling of shared memory was resolved for
distributed version control systems using a `Many Worlds' scope (see
for example \cite{git} for a review).  The essence of these approaches
is now part of distributed resource schedulers like
Kubernetes\cite{kubernetes} and message relays like Kafka\cite{kafka}.
We apply the same reasoning to construct an invariant but versioned
history over distributed changes for a more dynamic data service, i.e.
a generalized database, with contention free semantics.  We have to
solve the spacetime coordination issue, and select unique key names
for idempotently promised values.  We begin by defining the agents of
the system and their high-level promises.

\begin{figure*}[ht]
\begin{center}
\includegraphics[width=13cm]{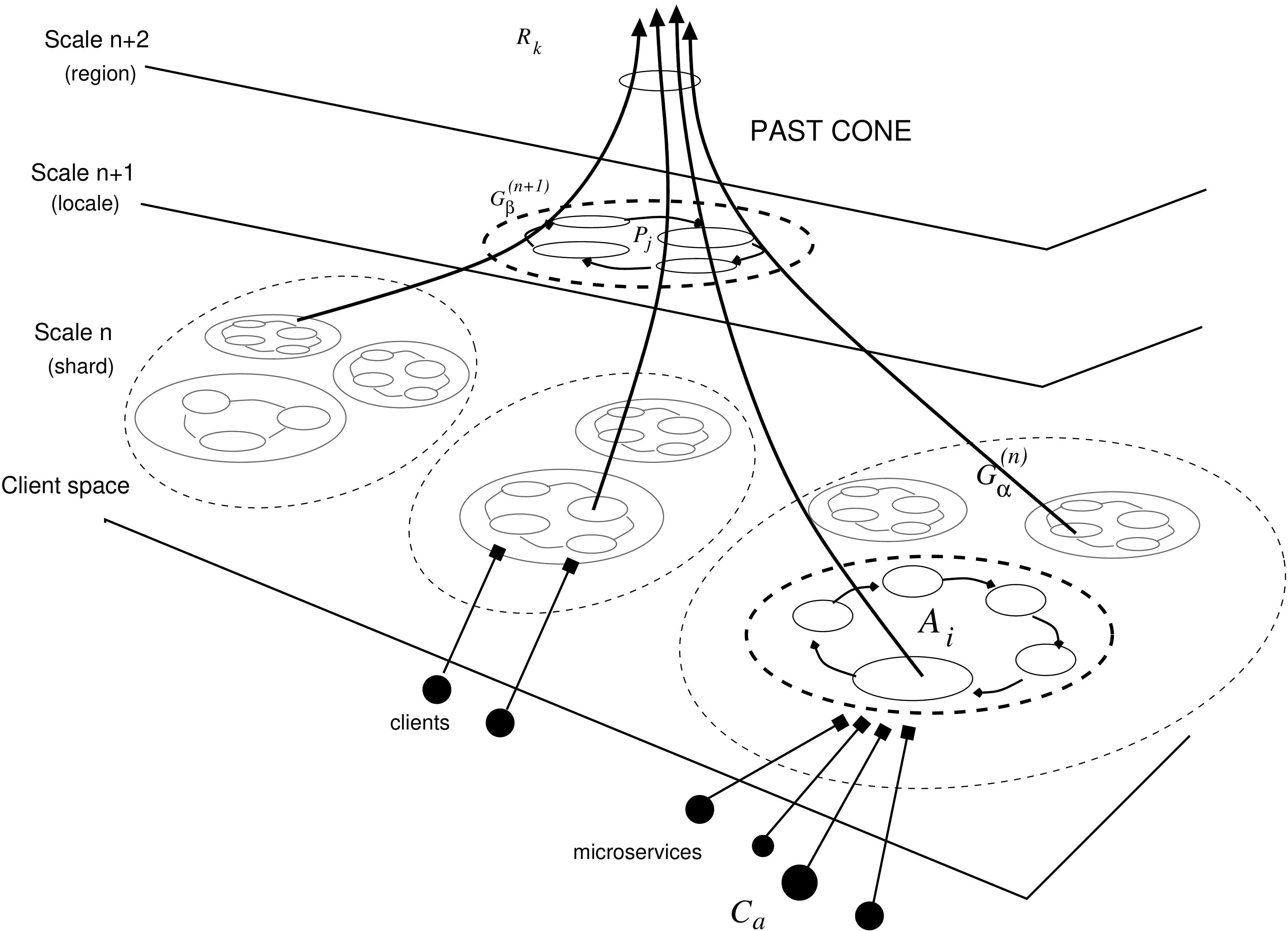}
\caption{\small The arrangement of agents in the configuration of a
  hierarchical service. The arrangement forms an implicit clock that
  computes the past cone of autonomous agents at different scales.
  The $A_i$ receive client requests independently.  Local requests can
  be completed independently; requests involving other shards may
  involve delegation. Committing of data is handled by parents and
  made public once accepted. Each local clock for $A_i$ is independent
  and unsynchronized, but if a search references data from other
  agents, then any agent can find out which versions belong in its
  past cone by consulting the hierarchy of parents as an index.\label{dbclock}}
\end{center}
\end{figure*}

Figure \ref{dbclock} shows the hierarchy of software or hardware
agents we refer to for the clock. We distinguish three roles for agents:
handler agents $A_i$, the parent handlers that aggregation handlers as children $P_j$,
and the root node(s) $R$. There may be several levels of parent handlers before
converging onto a master parent or root node.

The principal agents we shall refer to in our discussion
for a hierarchy according to their functional roles. We define them and their group labels
in this table:
\begin{center}
\begin{tabular}{c|c|c}
\sc Agent & \sc Role & \sc Cycles Group\\
\hline
$A_i$ & Handler agent& $G_\alpha^{(n)}$ \\
$P_j$ & Parent handler& $G_\beta^{(n+1)}(A)$ \\
$R_k$ & Root handler&
\end{tabular}
\end{center}
The arrangement of their roles is depicted in the figure \ref{dbclock}. This differs
slightly in detail from the idea in \cite{andras}.
Note in particular how spacetime is covered by a spanning tree composed from cells, each of which
contain cyclic `ring' structures.

The basic agent $A_i$ could be
associated with a server process running on a computer, but this
association is increasingly speculative given the many layers of
virtualization involved. The $A_i$ form collaborative groups, whose
roles are elaborated below. Notably, the members of a group form a
ring which becomes a tool for interleaving activity within the clock.
These groups a joined together by parent groups, also formed from
basic agents like $A_i$ but we call these $P_j$ for clarity, with
a different role to play. The parents, in turn, are joined together by
parents of parents, and so forth.

To avoid an excess of notation, we only refer to a single bundle of
threads from agents $A_i$ in a single cell, via a single parent, up to
a single root node. $G$ refers to a single cell or group of handlers $A_i$ with
a common parent $P$ in what follows.

The basic role of the $A_i$ is to be an independent transaction
processor, with its own private storage. We can think of these as
shards with transparent replication for redundancy, but a very similar
configuration could also be used for replication using the same clock
approach. The latter would be similar to the use of an entanglement
approach\cite{paulentangle}.  This means the role of a group $G(A)$
formed from the $A_i$ is as a database for `local concerns'.  The
precise meaning of this is ambiguous in a networked world.  The basic
role of the parent nodes is to be a single point of calibration for
the clocks of independent groups $G(A)$. The same applies across the subsequent levels
by induction.

\section{Design promises}

\subsection{Strategy overview: a global data pipeline and clock}

\begin{figure}[ht]
\begin{center}
\includegraphics[width=6cm]{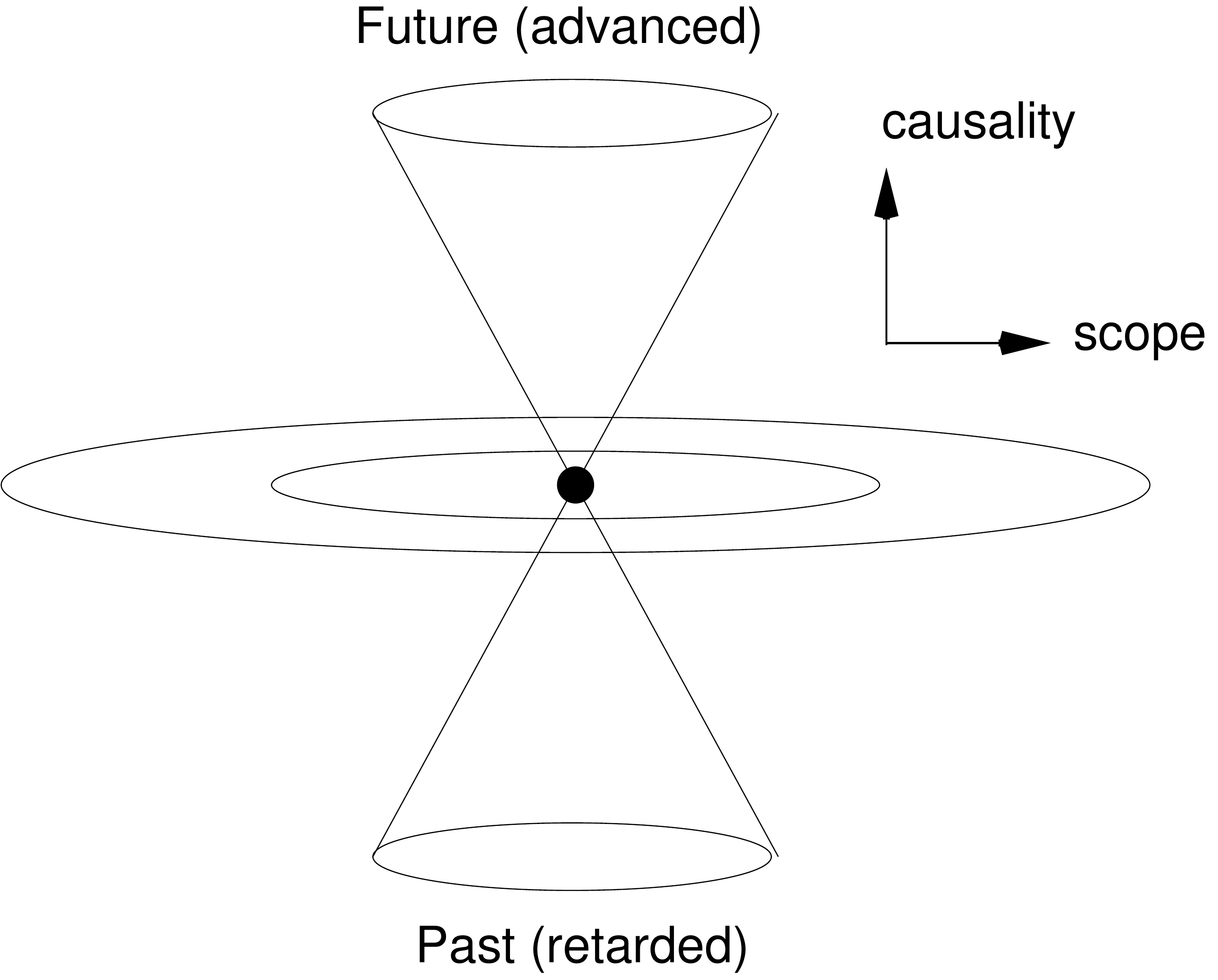}
\caption{\small The meaning of space and time in a data system: time
  is change, and space is independent context (memory or state).  The past and future
  cones that define the spacetime of each process. The past refers to
  data already archived immutably in the proper timeline of a process.
  Transactions at the meeting point can only depend on past key-value
  pairs (retarded boundary conditions). The future of possible states
  generated by the `now' point depends only on values from the past
  cone, not from anywhere outside. Deferred outcome processes may
  depend on future events implicitly (advanced boundary conditions).
  When these overlap, we need to use process context to disambiguate
  semantics.\label{cones}}
\end{center}
\end{figure}

Before turning to details, let's summarize the promises our
hypothetical distributed data service would be expected to keep:

\begin{enumerate}

\item Service handlers promise to accept client transactions that
  involve storing and retrieving versioned data histories, destined
  for one or more data subscription channels, each with different
  policy-defined semantics (suitable for different applications). The
  default channel always shows data service snapshots with `latest official
  version' semantics, for all key-value pairs $(k,v)$.

\item Each channel promises a consistent ordered sequence of versions, forming
  a causal history of earlier written $(k,v)$ data.

\item Each handler promises to expose client read-queries only to data
  values that are fully resolved and invariant, from the past time
  cone the query concerned (see figure \ref{cones}).

\item The total system promises to capture and preserve the {\em intended total order} of all
  serially executed changes to a given key-value stream by a given
  client, and to define the {\em partial order} of all causally
  unrelated (parallel) queries, according to a fair policy of
  interleaving.
\end{enumerate}

A rough schematic of the algorithm is as follows:
\begin{itemize}
\item When any command or transactional procedure is started, the cone of
absolute past for the process is determined from the global clock
infrastructure. This defines what shared values the process can employ
for its computation.

\item Once a transaction procedure is running, it may accumulate temporary
writes.  Temporary writes, such as those used in the
calculation of sums and composites etc, will never be seen by any other
process. 
\item All writes, whether intended to be permanent or not, are written immediately to
the private workspace belonging to the transaction, in the manner of
private local variables. These are never `committed' or published. 
\item In order for written data to become permanent new versions of
  $(k,v)$, i.e. invariants of the service, a winning representative must be selected from
  possibly any competing contenders, and be authorized by adding it
  to the input pipeline of the parent handler, as in figure
  \ref{dbclock}.
\end{itemize}
This distributed data pipeline actually forms a `global clock' to batch data
into a coarse grained partial order, by virtue of its policy
for interleaving authorized `commits' from all cells.\cite{andras,koalja}.
The result of this schematic process is to curate a set of consistent channels
(including the default `latest version' view). The 
handling agents $A_i$ promise to ensure that no two processes are ever allowed to
`commit' a write to the same key in the same global time slot;
however, alternatives and race losers could remain in semantically
separate channels.

A key promise for data subscription replicas is that all
agents at the same spacetime coordinates will see the same
past data cone. We need to define sameness:
\begin{definition}[Sameness of snapshots]
  Two snapshots may be called equivalent (or `the same') if every
  key-value pair in their past cones is identical. This occurs if the
  two snapshots are computed from the same global time coordinate (on
  the hierarchical clock).
\end{definition}
Agent changes are clearly fluid and changes are continuous, so the
likelihood of agents seeing precisely the same information for very
long is small if changes are frequent. The key point is that this is
deterministically possible, so that every agent is part of a single
shared process (i.e. in the same branch, in version control parlance).

\subsection{Locality as a design principle for correctness}

The challenge of scaling data is usually presented as the abilty to
cope with a large amount of it. The harder problem is to ensure that
all the promises and meanings we intuit about data (on a human scale)
can be maintained as services are scaled up and down. The main method
at our disposal, for preserving these semantics, is to trade the
ability to see change for greater dependability.

\begin{definition}[Service correctness]
  Correctness of a service is defined to be the state in which
  explicit promises have been given for the service, in each
  distinguishable context, and where these are kept for all observable
  outcomes.
\end{definition}
Delaying the observability of data affects the `liveness' that we can
promise about changes, so we need to ensure low latency responses on a
number of levels.  Writes can be effectively immediate (lock free), but 
the visibilty of the changes for others will depend on several factors.
An equitable solution, with easily understood
semantics is to use locality as a principle (i.e. what we experience about
ourselves is local and familiar; what others experience far away is
less visible and less relevant to us). When we need to read data, the larger
the spacetime region involved in collating the data, the greater the range of
uncertainty in its liveness, partly due to the time it takes to retrieve the
past cone for a transaction, and partly due to the time it takes to integrate
changes to the cone into the coordinate system, as new data are written.

{\em Locality} is the principle by which the observation and capture
of changes are kept as close to the point of interaction as possible
for speed and accuracy. Other issues, like energy conservation are
also benefitted by local thinking. Data searches, however, span data
that are distributed over multiple shards and possibly stored over a
wide area.  Locality is particularly important for calibrating changes
(writes) accurately in time, since a single point of change makes
a single version semantics straightforward.  For read-only interactions,
locality is not as much of an issue as long as records are proper
invariants (sometimes called immutable values) of spacetime processes.
However, variations in version over spaced caches is a familiar problem
where data are shared and dynamic.

Traditionally, one applies push-based thinking to build services.
There we control the ability to expose or limit access to resources
using mutual exclusion (e.g. mutex locks). These serialize access for
critical sections and impose a local order, which is familiar, but
costly in terms of delays and busy waiting.  Version control schemes
for parallel workflow, and Content Free Replication Types (CRDTs),
handle this using private branches. We can combine branches for wait-free
pipelining with restricted observability for avoiding contention. This
is analogous to `look ahead' and deferred
publication\footnote{Deferred observability leads to change phenomena
  analogous to non-local quantum probabilities, presumably for the
  same reason: observation is decoupled from the processes that
  marshall the configuration of resources.}.
Once data are invariant and uniquely referencable, published data can
be replicated any number of times, and temporary caches can be brought
arbitrarily close to clients for efficiency using some state
replication policy\cite{helland1,helland2}.

\begin{figure}[ht]
\begin{center}
\includegraphics[width=7cm]{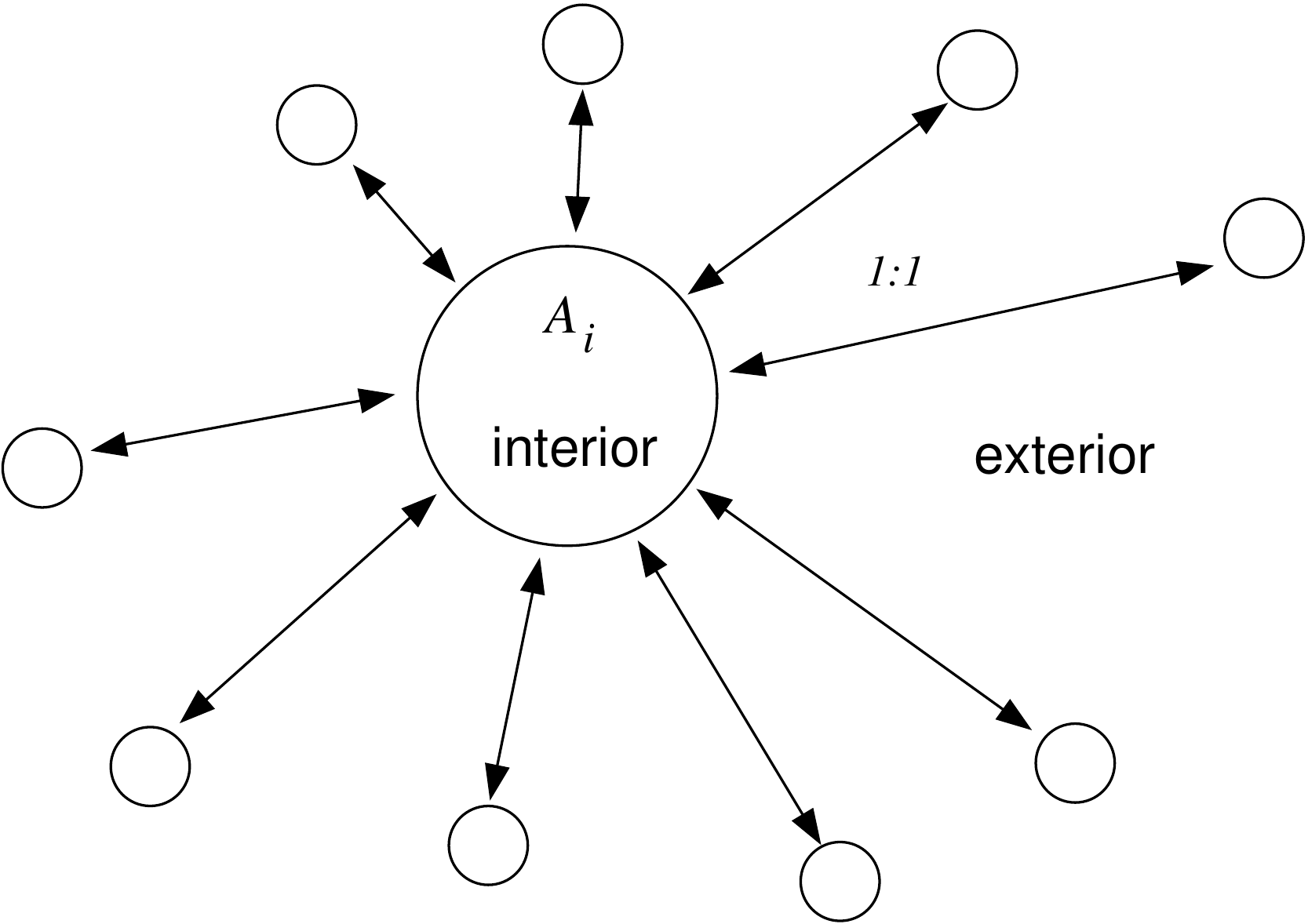}
\caption{\small One to one correspondence with a singular source leads to a single valued
function of inputs. Thus logical centralization and single-valuedness are related.\label{intext2}}
\end{center}
\end{figure}

Locality is closely associated with {\em centralization} (see figure
\ref{intext2}), i.e. the use of a single point of change---though the
latter is often represented as a controlling concept rather than a
calibration issue.  The main functional meaning of such localized
centralization is to create single-valued outcome, by using a single
process for calibration\cite{treatise2}.  The Downstream Principle in
Promise Theory\cite{treatise2} implies that time calibration (and
ordering) is arbitrated locally.  So, the strategy for ensuring
consistency would be to localize all shards of shared data to a small
area.

\begin{principle}[Locality determines correctness]
  For each promise kept by a service handler agent at a single
  location, there should be a corresponding promise, with the same
  semantic structure, kept by the entire system, but with an adapted
  set of dynamical parameters.
\end{principle}

\section{Causality}

At every location throughout a system, causal changes originate from
the interface between handlers and clients (what we call the `edge' of
the system).  We expect an intended change to be reflected in what we
see from that moment forwards. We expect what we already know to be
dependable, subject to some possible interference by competing
clients.  Our local region is where causal `intent' is sourced from,
and thus we consider this to be the origin of meaning for data---close
to the context in which changes arose.  Locality thus
tells us there are two issues in resolving this problem:
\begin{itemize}
\item Resolving intentional order for versions of $(k,v)$ at each handler location $A_i(k)$, and

\item Curating a cooperative non-local interpretation, i.e. one that spans a broad reservoir of data
that spans a wide area of spacetime.
\end{itemize}
These are the challenges we describe next, by designing a pipeline for
spatial integration with temporal flow. On a theoretical level, this
involves some deep issues about the exchange of interior and exterior
time, by redrawing boundaries between public and private scope, when
scaling\cite{virtualmotion,spacetime2}. Naturally, there is no free
lunch when it comes to consistency. We can eliminate input/output
bottlenecks, but only at the cost of a delay in the availability of
consistent information (extra latency in computing parlance).
Pragmatically, we need to design a pipeline to publish data changes
using a single-valued coordinate system, based on past, present, and
future.

A part of causality, which is not often included in academic
presentations, is {\em context}.  The purpose of data updates may
be of significant importance to the behaviour of a service.  This is
one reason why we need so many different kinds of database for
different applications.  The differences are particularly noticeable in
replication of state. For example:
\begin{itemize}
\item When incrementing a counter or balance between many clients (coordination).
The value corresponding to a primary key is being overwritten continuously
and there is a `current' or `latest' value.

\item When collecting a data lake or warehouse. Data are cached and never or rarely overwritten.

\item When writing a timeseries, journal, or log. Data are appended but never overwritten.

\end{itemize}

\section{Continuous Integration Coordinate System}

In this section we detail the semantics and dynamics of the spacetime
coordinate system used to label committed data.  We understand data
processing `transactions' to be extended procedural threads, not
merely single read or write events. We assume that a client $C_a$ is
the broker responsible for managing a transaction. It must therefore
have access to the time counters of the hierarchy. We shall not
discuss possible implementations of these agents.

A transaction may read many records from anywhere in the dataverse,
combine them, and write a single result or update every record one by
one. Such large scale operations lead to extended contention, which
interferes with other operations.  These explosive cascading changes
are hard to understand, for humans, as we tend to think in terms of
our own narrow field and perception of time.  Time consistency is
built around the concept of single cause leads to single response (1:1
causation, see figure \ref{intext2}), so it would be advantageous if
technology treated every such transaction as if it were an
instantaneous operation too. We can manufacture this illusion by
restricting observability, in the same way that human experience is
limited.

\subsection{Localization of commands and transactions}

\begin{figure}[ht]
\begin{center}
\includegraphics[width=9cm]{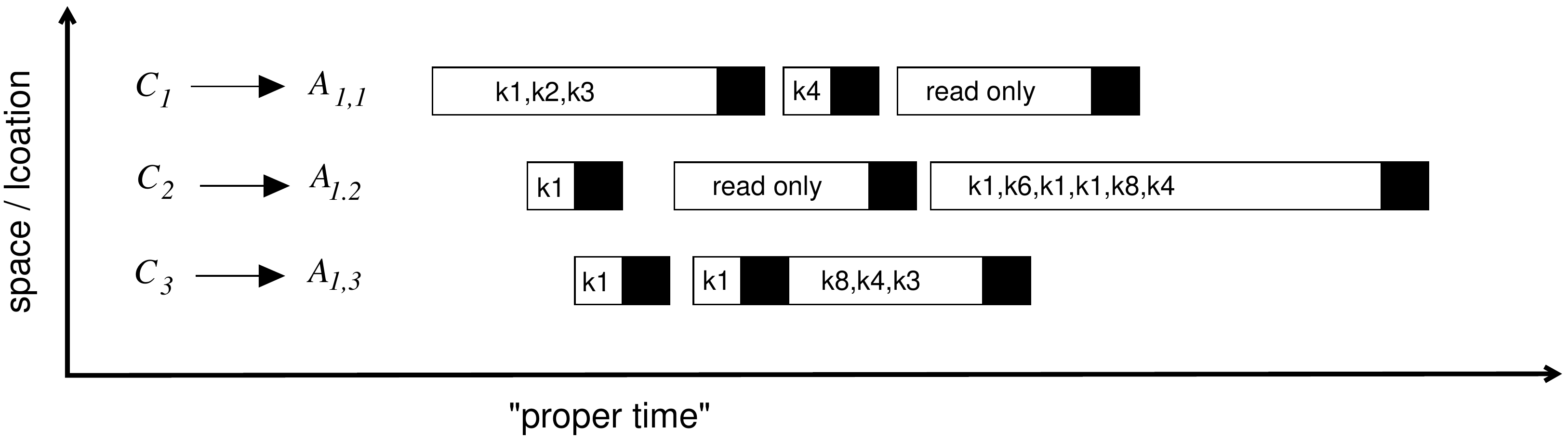}
\caption{\small Data transactions with a number of service agents $A_i$, responsible for
reading and updating data for clients $C_a$. Each $A_i$ writes data only for keys that it owns,
but may need to read data from other agents around the world. 
The dark regions indicate potential `commit' operations at which the sum effect of
a transaction is recorded. Without assigning meaning to these races, the value of shared key-value
pairs is ambiguous. Only a policy, which settles the intention behind the updates, can resolve the
ambiguity. One has the choice to keep all writes in separate channels or to merge them into a
common view by some policy enacted by the handler.
\label{transactions}}
\end{center}
\end{figure}

Consider a number of clients $C_a$ interacting with a stateful
software service---a database---through a number of parallel handlers
$A_i$ (see figure \ref{transactions}).  Each stream of changes from a
clients $C_a$ to a handler $A_i$ forms a converging data pipeline, with
possible contention around changing a notion of `current value' (see figure \ref{transactions}).
Each interaction thread between $C_a$ and a single state location $(k,v)$ at $A_i$ forms its
own proper time history unless they collide with one another at a particular location $(k,v)$
during the same causally dependent interval.

Since the key names are shared between all the competing threads,
indiscriminate updates will then interfere destructively with one
another unless we intervene to keep them apart. Without defining
explicit semantics for a resolution, this would lead to an unclear
outcome. Whose intended change wins this race?  The underlying problem
with this view is that the concept of `current value' is neither an
invariant nor uniquely definable property of a distributed system:
it's an illusion of a single observer's timeline. Different clients
understand `here and now' differently from others.  The curator of a
shared outcome has only two main options for taking what was stacked
in parallel and mapping it into a serial stream by multiplexing space
and time:
\begin{itemize}
\item {\bf Time separation}: Agents race and coordinate individually
  to overwrite the latest version of a key-value pair. Coordination
  must be maintained between the agents using some form of
  interprocess semaphores or mutex locks.  All possible futures
  `collapse' into a single shared version of global state. In case of
  contention, `rollback' of the shared state may be required to
  maintain the illusion of correctness for a single timeline; however,
  the fact that there is an interference of more than one agent's
  intent makes rollback of client state impossible (non
  single-valued), so attempting rollback is fundamentally ambiguous
  for clients (see \cite{jan60} for a discussion in relation to division by zero).

\item {\bf Space separation}. Alternatively, each agent works in
  separate private branches of the data namespace they wish to change
  (as in distributed version control\cite{git}).  Coordination is
  unnecessary, because the branches do not interact except through a
  shared past. This is the `Many Worlds' version of global state. It's
  well known in programming as the local private variable approach.
  Future merging of these separate branches into a permanent shared
  result is a policy-based decision and requires additional work (no
  free lunch).  The receiver is free to curate any combination of
  changes, or even invent something totally new, in
  principle\cite{koalja}. Rollback is never needed in this approach as
  every change can be considered intentional. Unused branches can be
  garbage collected after a certain horizon to avoid an exponential
  explosion of state.
\end{itemize}
We adopt this latter approach, where the correctness of a merger is an
{\em ad hoc} decision for the receiver of the two branches, and we
look for a resolution that's equitable in a general setting. 

It's important to realize that the original client sources of the data
have no access or `right' to resolve conflicts downstream after they
have handed over data to the receiver.  Moreover, the perceived order
in which changes are made across different branches is not necessarily
a relevant criterion for choosing alternative outcomes, since it's
only observable by the receiver as a single point of
calibration\cite{promisebook}.  We compromise between these methods to
curate a fair process of interleaving, performed in `real time'. This
approach is sometimes known as Continuous Integration in software
engineering.

\begin{remark}[Data pipelines]
Data pipelines have a exactly analogous problem to solve: data from multiple sources,
e.g. different sensors, arrive at some aggregator process and have to be
combined (say into a statistical overview). Coordination is needed to decide
which sampling timeframe data belong to. The receiver is the only agent
capable of deciding this to create single-valued result. Policies for
integration were described in \cite{koalja}.
\end{remark}

\subsection{Scoping of program variables for data transactions}

The declarative aspect of data query languages presents a scoping
challenge.  Functional and imperative programming languages solve the
problem of transactional integrity for procedures using private local
copies of data to build `pure functions'.  Inputs are copied into
local memory by value and the function then works on the copy, leaving
the original immutable. Several functions can operate on their own
copies independently, without interference.  Databases have yet to
fully take such methods on board, though snapshot isolation is a
partial adoption of the idea\cite{snapshot1,snapshot2,snapshot3}, and
CRDTs are the natural correspondence\cite{crdt}\footnote{The Datomic
  database, for example, has drawn this comparison with the
  immutability of past events.  Hickey\cite{hickey1,arewethereyet} has
  popularized this in talks, though the formal details do not seem to
  have been published.}.  We apply this idea to the formation of
channels for named data distribution.

\begin{definition}[Private proper time or version]
  Let $(k,v)$ be a key-value pair owned by a handler $A_i$.  A private
  proper time increment or {\em version} or version of $v$ (which we
  denote $v_n$ for version $n$, where $n$ increments with $t(A(k))$), is promised by the handler thread at
  $A_i$ on the arrival of any new transaction, for any $(k,v)$
  accepted by $A_i$ from any client.
\end{definition}
We need to publish an outcome, from a single winning update, in a sequence at each
point $A_i(k)$, at the moment changes are made public, thus:
\begin{definition}[Public (shared) proper time or version]
  Let $(k,v)$ be a key-value pair owned by a handler $A_i$.  A public
  proper time increment or {\em version} or version of $v$, is promised
  to occur on committing the outcome of a winning transaction for all $(k,v)$
  altered by the unique client thread. A public version number has the role of a timestamp,
determined by the top root note $t(R)$.
\end{definition}
Notice how the decision to see and accept an external write command is
an autonomous decision of the handler $A_i(k)$. This makes the definition of any policy
a promise that only $A_i(k)$ can make, for each $k$. 

\subsection{Proper time simultaneity}

We can now define the meaning of `at the same time' with respect to
proper time. This is different for reading, writing, and committing.
Note that a client $C_a$, which brokers connections with key
handlers, is different from a key handler $A_j$, which is only responsible for writes
(and perhaps reads) for its custodial keys.
\begin{definition}[Simultaneity of reads and writes]
Read and write operations are interior to their parent transactions.
If the coarse (exterior) time counter to two operations is equal in any counter 
component of the tuple $t(A),t(P),t(R)$, then they are simultaneous over the scale represented 
by the tuple component.
\end{definition}
Scaling this to extended collections, we have:
\begin{definition}[Simultaneity of commits]
  Two committed values are said to be locally simultaneous if their outcomes are accepted
by a parent in the same time interval tuple, $(t(P),t(R))$. 
\end{definition}
In practice, this means that we consider data values that belong to the
same proper time `snapshot' to be simultaneous. The subtlety lies in the
fact a snapshot of smaller spatial reach can be resolved more
accurately (with finer granularity) than a larger one.

We state now a principle of minimal scope, which is central to resolving a global notion of consistent state:

\begin{principle}[Minimal scope published by `commit']
  All pairs $(k,v)$ written by an agent $A_i$ are made as interior
  promises, whose outcomes are unobservable by exterior agents
  (clients, etc), until a decision is made to publish the result on
  the exterior to become part of a shared past cone. Only exterior
times are observable by client transactions.
\end{principle}
The implication of these definitions is that the `commit' operation,
for a database, is the decision to propagate the outcome (of a write) up
the hierarchy to the root node, adding to the observable values of a
subscribable channel.

\subsection{Semantics of the client interface}

We begin at the edge of the system, where several clients $C_a$
(i.e. the sources of data and work) contend for access to a single key $k$.
We'll assume the client itself is the broker for working with multiple shards.
A client can find a shard handler's address via a directory service (analogous to DNS
for the database), which is an invariant since the location of a key does not change.
Thus, all requests to read or write $k$ converge on a single data
handler agent $A_i$, whose task it is to accept and record changes in
the form of key-value pairs $(k,v)$:
\beq
C_a \imposition{+O} A_i,
\eeq
where $O$ is an operation (private read or write).
The reading of a known data value version may go to its
custodial handler $A_i(k)$, or to any read-only cache or replica.

In the language of Promise Theory, interactions with the handler begin with clients
$C_a$ imposing some command (+), resulting in a transaction $T_x$, onto the service
handler $A_i(k)$\footnote{In general, handlers might alternatively pull data from known
  sources by a regular schedule, e.g. when replicating data for
  backup, etc, but one normally thinks of data arrivals as randomly
  imposed events.}
This interface defines the edge of the system, with $A_i$ on its interior
and $C_a$ on its exterior.
In general, the index $a=1,2,\ldots$ may run over many different clients,
and the index $i=1,2,\ldots$ will run over many different handlers.
In order for a response to be triggered, the handling agent has to
accept some or all of these transactions
\beq
A_i \promise{-O'} C_a,
\eeq 
and the accepted (-) portion is $O \intersection O'$. This allows for access controls
and privilege levels, as well as throttling of commands from clients.

Interactions or transactions between clients and handlers are not
necessarily point-events. They can last for extended intervals,
involve many reads and writes, and thus span data that are distributed
across a wide area of system (in space or time). The control structures for
this lie with the client. The convergence of
more than one client onto a single handler (see figure
\ref{transactions}) may therefore involve significant contention for
the same resources, especially for long running transactions. 

The handler multiplexes parallel threads and thus
entangles clients' outcomes together. Where client dependencies
overlap, this leads to a `race' to acquire exclusive rights to change
states $(k,v)$. The handler has to resolve these races fairly and
quickly to ensure that the data service keeps its promise of being a
single-valued function of $k$ for all future interactions, as well as
a timely reflection of current events---this is the mathematical
expression of what's expressed as {\em data consistency} in the
technology literature.

\subsection{Client race adjudication and policy semantics}

Resolving the outcome of races to change a single key involves the following criteria:
\begin{enumerate}
\item The ability to retain and separate past $(k,v)$ values, i.e. data histories, whose values have
  already been determined in the immutable past, from values that are in dispute
  in the present or the future. We call this the determination of the
  {\em past data cone} (see figure \ref{cones}).

\item The ability to identify collisions (sooner rather than later) when the evaluation of commands or transactions
will contend for certain keys. This may be predictable
from the semantics of the command, but---in the worst case---processes may have to play out until the last moment
of a collisions in order to determine which keys will be affected. 
If the process is long, cancellation of a particular client's transaction may involve a significant waste
of time, energy, and other resources.
\end{enumerate}

The semantics of resolving races may impact data flow significantly. A millisecond
difference in latency could be the difference between a detectable collision or none.
\begin{itemize}
\item Should we insist that all but one of a set of parallel transactions fail in case of a collision?

\item Should changes by queued with a definite order?

\item Is it right to allow a value, which was just written, to be overwritten by another process
just because it narrowly avoided collision? Should key-value pairs have dead times, as neurons
do to prevent thrashing?
\end{itemize}
These questions are rarely asked of services, but they express
particularly relevant aspects of intentional client behaviour.  Was a
particular update a characterization of the moment, whose relevance
was temporary, so that a single failure could be dropped (unreliable
delivery)? Alternatively, was it a critical step in balancing some
account of a broader process whose loss could have enormous impact?
Such a characterizations have to be embodied by a policy for resolving
missed opportunities to write data.

Simple policies for selecting a winning transaction are well known
from scheduling, e.g. First Come First Served (FCFS), Shortest Job
First (SJF), etc.
The mode of resolution could depend on the visibility of the outcome:
will the result of a write be shared between all future clients, or
will it be private for certain groups? Groups can separate flows
into `channels', implying a policy for the eventual channel a
transaction will end up in, i.e. will transactions collide or be
placed in separate branches?

\subsection{Policy for resolving write collisions}

Consider then how parallel changes are handled when contending to change a single key.
Collisions between clients occur when at least one of a number of simultaneous transactions
attempts to write the same keys that are being read or written by others
in parallel.

No interior writes are visible outside the transaction, for its
duration. Commit operations are all registered at the end, as
exterior time $\tau$, i.e. at the close of a transaction (regardless of
when they were written in interior time).  Concerning the interior
time, if a value has been read at $t_21(A(k))$ by $T_1$ and it is
written later at $t_22(A(k))$ by $T_2$ then then there is a collision
and we need a policy to resolve. If the policy is FCFS:
\begin{itemize}
\item $T_2$ will never see $T_1$'s changes, since they lie in its
  future cone. So $T_2$ cannot perform any reads or writes, to
a key affected by $T_1$, until after $T_1$ has committed.
  Increments must remain relative until the moment of committing (i.e.
  the values should not be based on an earlier read).
\item If $T_1$ and $T_2$ both read and write the same key, e.g. both
  increment or decrement, then $T_2$ must be rejected by the handler,
  because we know that the value it sees is going to be invalid by the
  time it might commit its own changes.
\end{itemize}
We point out that the contextual semantics of the use-case is
important here.  When the transactions involve relative updates, they
have to read the previous version in order to compute the new version
to write, so overlaps are potentially complex constructions.  As
mentioned above, we won't discuss the implementation of these
promises, which are challenging.

One could try to make the algorithm perform some complex traffic management,
but this could lead to presumptuous behaviour. We have to question why two agents
would be allowed to alter data simultaneously in the first place. This is a question
to be resolved by the imposers of intent. Promise Theory tells us that impositions
can't be resolved by the receiver, and are therefore likely ineffective in
generating their intended outcome. Some out-of-band communication between
clients is thus likely required to find the `correct' resolution to this contention.
A simple failure of an overlapping transaction may delay a client, but is a
neutral resolution.

With this meta-policy, there are only two kinds of collisions to handle, which correspond to $\pm$ promises
in the data interactions:
\begin{itemize}
\item {\bf Collisions of intent (+,write)}. One or more client transactions attempt to claim the same version $(k,v)$,
by writing to it in the same proper time interval.

\item {\bf Collisions of dependency (-,read)}. A client intends to write $(k,v)$ and other parallel processes depend on the
value of $(k,v)$ for their own outcome.
\end{itemize}

\begin{proposition}[Read conflicts] 
Reads can be avoided by promising
  no observability or inclusion of data values committed at or after
  the moment at which a transaction began, in a later version.
\end{proposition}
In other words, transactions may only be granted access to read from the {\em past cone}
in a transaction evaluation (see figure \ref{cones}). 

The selection of which version from the past to read remains the
responsibility of clients $C_a$, by the downstream principle. The
common and natural choice is to read latest version at the `top',
`end*, or `head' of the chain of values $(k,v_n)$ where $n \le \text{now}$.

\begin{proof}[Proof of absence of read conflicts]
  If all changes to data are assigned a greater version number in each
  channel, both privately and publicly, then there can be no version overlap
  between $r(v(c))$ and $w(v'(c))$, provided
\beq
A_i \promise{v'(c) > v(c)} C_a,
\eeq
which is a promise of the handler.
\end{proof}
These policies thus resolves the promise to avoid read conflicts---a promise rather than
a guarantee, because it assumes the handler agent keeps its promise. The presence of
bugs or other flaws can still undermine this.
\begin{figure}[ht]
\begin{center}
\includegraphics[width=8.5cm]{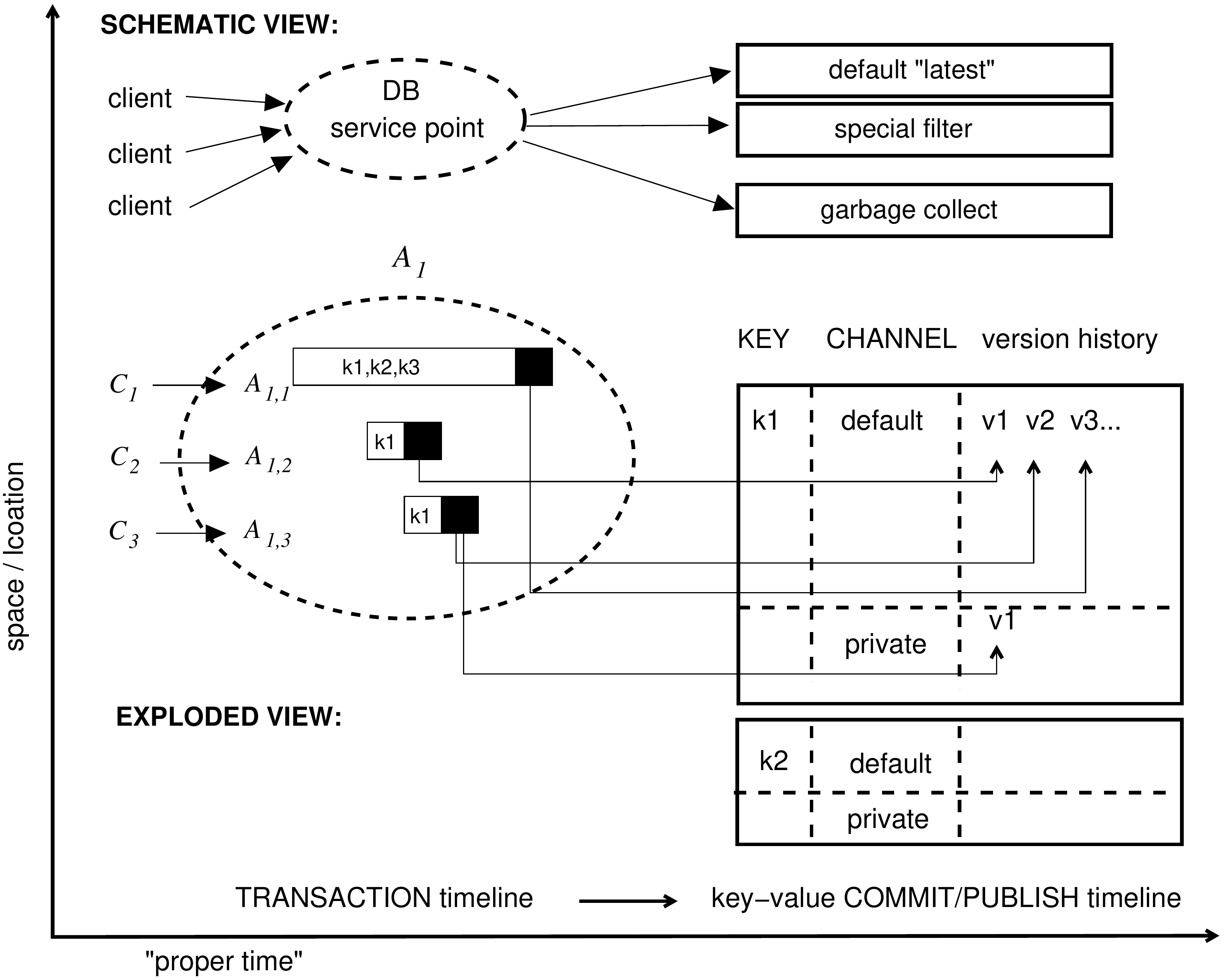}
\caption{\small Transaction race resolution is performed at the edge,
  by authoritative key handler agents $A_i$. Clients, from any
  geo-location, initiate processes in a data handler, which may handle
  multiple sessions in parallel. A handler assigned to a client
  process will only read from the past cone measured from the starting
  time of the request. Any changes to values are written `immediately'
  by handlers into private workspace, and the final publishable set of
  changes is committed in a single clock-tick at the end of the
  transaction as its final act (shown as dark squares).  Overlapping
  processes race one another relative to the final channel curator
  process.  The channel curator keeps time by a hierarchical
  distributed clock, and accepts convergently named slot-reservations
  for a single key/channel on a FCFS basis, and places them into a
  timeslot within its publication schedule from its current future
  cone.\label{channels}}
\end{center}
\end{figure}

We can thus turn to writes (see figure \ref{channels}):
\begin{proposition}[Write conflicts] Write conflicts can be avoided by:
(i) immediately and serially performing writes to a namespace
which is private to the transaction, and (ii) selecting
only one winning client from the list of clients that would commit (publish)
their changes in the same time interval.\label{propw}
\end{proposition}
The freedom to pursue variants of this lies in the ability to curate different channels
from the stream of accepted temporary values.
We define the default policy by
\begin{definition}[Default collision policy (FCFS)]
The first transaction to be accepted from $A_i$ queue will always have the right to
commit all values in its write set $w(C_a)$. Any later transaction whose write $w(C_b)$ set overlaps
with the first is immediately terminated. If the two processes write to different channels,
there is no collision.
\beq
A_i \promise{\text{fail/terminate}\;|\; w(C_a,(k,v)) \intersection w(C_b)} C_a
\eeq
\end{definition}

\begin{proof}[Proof of absence of write conflicts]
  By writing serially there will be only one causally obtained value
  of each $k,v$ at the end of the transaction, which remains
  unobservable except by the handler.  The handler can promise that no
  values written by some other process, between the start and the end
  of a given transaction,  could have been seen by random selections of clients,
  since all such values are unobservable until after resolution. Thus
  we need to show that all values made public are fairly available to
  any client that attempts to read them, after the transaction has
  completed, and that no other transaction started since could have
  altered this sequence.
\end{proof}
Data that need to be written by a command are written immediately into
the channel's private namespace, and thus remain unobservable
unless promoted to one or more other channels, e.g. default channel,
debug channel, etc. by a successful FCFS completion, or purged after all possible
use has expired.

\subsection{Avoiding indeterminacy due to push based semantics}

In Computer Science, push-based update events (`impositions' in
Promise Theory language) are almost universally adopted as the model
for signalling changes.  With push methods, agents who {\em intend} a
change implicitly assume that their intent will be immediately
respected in the final outcome. In other words, they can bank each
change and proceed with impunity. It's hard to come back from walking
out on that ledge if others are using the same ledge at the same time.
Koalja adopted a message passing scheme of queued notifications along
with a separate pull-based publish-subscribe data
channel\cite{koalja}, showing how to scale data pipelines with
decoupled signalling and data for greater reliability and scalability.

To define a service as a single-valued total function of its inputs,
for all times, we need to give meaning to what can be observed by
clients and handlers on both the interior (within a private processing
context) and on the exterior (shared as a final outcome).  Engineering a
form of determinism plays a key role here.  This is a particular
challenge when interactions with the clients may be either short or
long in duration relative to the rate of updates.
The downstream principle in Promise Theory, which represents local
causality, on the other hand, tells us that it's the recipient service
that determines the single-valuedness of the outcome---not the
client \cite{treatise2,promisebook}---and therefore it has to be both the calibrator and arbitrator
of ambiguities for such decisions. This includes those that occur in
client collisions. The goal is clearly to
make such a service handler respect the intentions of the client as much as
possible.

\begin{figure}[ht]
\begin{center}
\includegraphics[width=8cm]{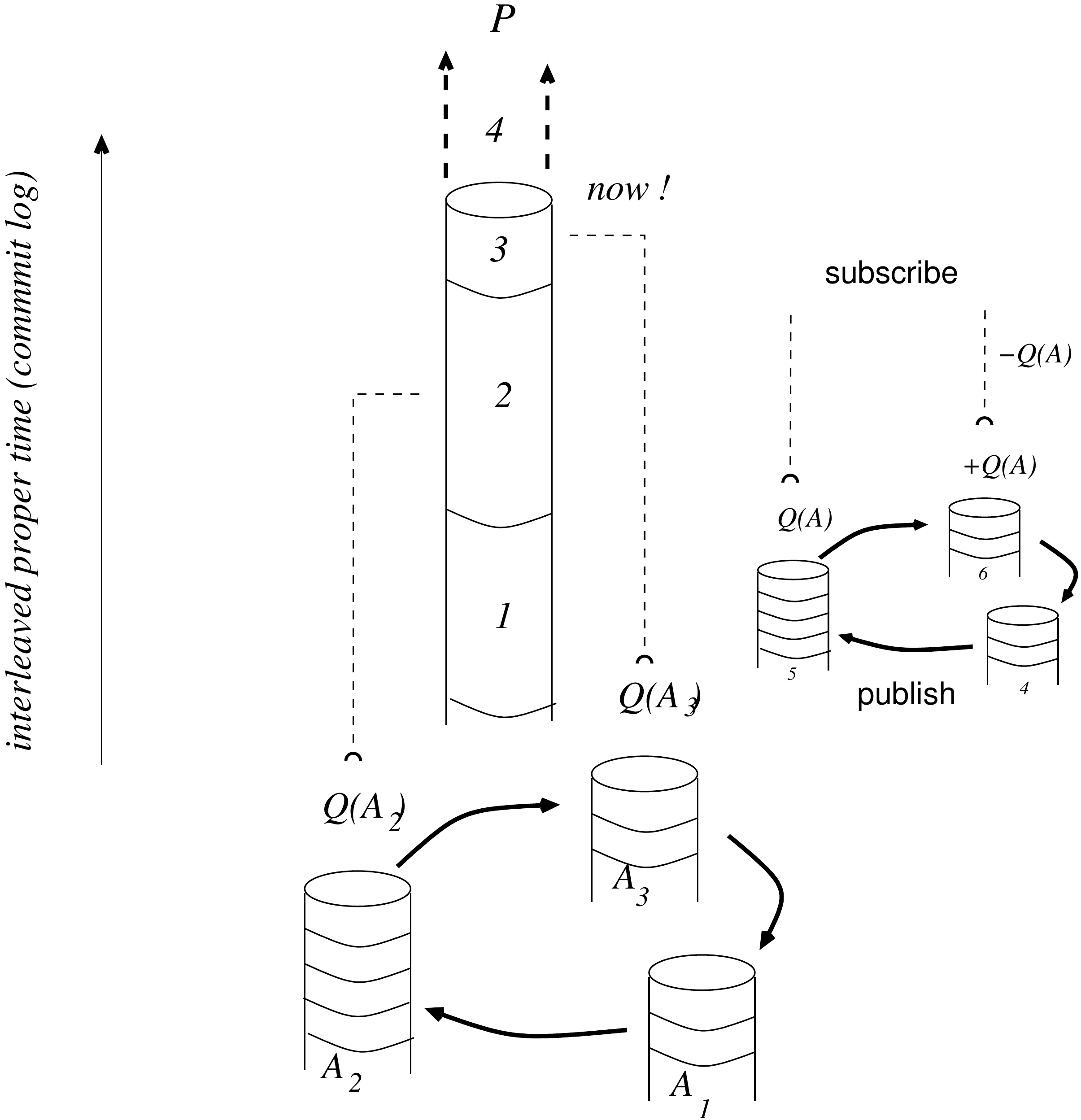}
\caption{\small Data from parallel handlers can be interleaved for
  indexing.  This also allows parents to create partial order based on
  coarse grained time intervals, by polling the client queues each
  time the client queue holds the token for sharing. It records pairs
  of token counters at handler and parent $t(A_i),t(P_j)$, thus
  allowing data from different $A_i$ with the same $P_i$ to be
  partially ordered with respect to $P_j$.\label{interleave}}
\end{center}
\end{figure}

In our model, the automated integration pipeline, which spans all
clients and data handlers, assigns a globally ordered version number
to all new data, using the pull-semantics to achieve several benefits.
First, the weak coupling of `pull' encourages fault tolerance and is a
natural choke for resolving traffic bursts with single-values
semantics. Second, it remains independent and can act {\em
  pre-emptively} as a scheduler. This prevents any hanging hosts from
bringing down updates from parallel handlers, and allows auto-recovery
(self-healing \cite{burgessC11}).  Finally, it acts as a calibrator for fair
interleaving (as in figure \ref{interleave})\cite{burgessrpc,remi3},
promising a consistent view of temporal order for all clients at any
scale of the hierarchy.

The proper timeline of an interaction is as follows:
\begin{enumerate}
\item A client $C_a$ pushes its intention to update  $(k,v)$.
\item If the handler $A_i$ for $(k,v)$ accepts the write, it selects a
  uniquely-ordered collision-free context namespace for temporary
  storage.
\item Accepted writes are locally wait-free, effectively immediate at
  the handler, but the total wait time for a client transaction with
  multiple writes is finite and is proportional to some monotonic
  function of the number of keys involved.

\item There is no impediment to reading from the past cone as long as
  there is service availability (including caches and replicas). The
  time to read key values is also proportional to some monotonic
  function of the number of keys involved.

\item Accepted writes are placed in the next timeslot of the handler's
  queues, where they will be interleaved into the timeline by the
  parent handler, and passed up the hierarchy at each scale until
  globally ordered at the root. While this is taking place,
  replication of the independently labelled writes can proceed in
  parallel, since lower key-value layers do not cache or depend on version data
  from upper parental layers.

\item Once indexed by the parent hierarchy, the new versions of the
  transactions updates become available all within a single timeslot
  of the global clock (i.e. simultaneously as far as the time cone of
  any new process is concerned).

\item Clients can then subscribe to any channels as soon as data have
  moved far enough up the hierarchy to be indexed and thus visible for
  subsequent reads.
\end{enumerate}

\noindent In broader terms:
\begin{itemize}
\item A policy decides which channels $c$ can be written to by the client.

\item Each client writes or overwrites temporary key value pairs $(k,v)$ within its own private space during the
  handler process immediately and with impunity. The map
\beq
(k,v,c) \mapsto (C_a,k,v,c)
\eeq
is collision free, where $C_a$ is used here to represent the transaction.

\item The handler of each $(C_a,k,v,c)$ looks for collisions, i.e. determines whether more than
one $C_a$ intends to change $(k,v,c)$ in a scheduled handler time interval.
If so, it selects a winner and returns with a refusal to accept the imposed
write to the client.
\item The surviving values are queued up by $A_i(k)$ into a common buffer $Q(A)$, which is
then emptied periodically by the parent pipeline receiver.
\end{itemize}
What happens after this is a matter for policy. Rejected clients could
shoulder the responsibility to retry
writing the value at a later time, or the handler could do this on their behalf. Neither possibility resolves the
meaning of having two clients attempting to change the same
value in the same approximate time-frame, but it resolves the actual collision. The matter of whether two
clients should be allowed to change the same $(k,v)$ is typically a matter for
access control policy.  In CFEngine, attempts to change a value were rate limited with deadlocks,
preventing thrashing as an acknowledgment of the spacetime aspect of intentional change\cite{lisa97113}.

For example, incrementing or decrementing a shared value is
commutative and deterministic and easily understood, but overwriting
an absolute value with an unrelated value is not commutative and would
effectively turn $(k,v)$ it into a random variable.  This is not
something a service can resolve except by denying access.

A global pipeline policy for enforcing a single-valued coordinate-history at each
local agent simplifies and potentially speeds up the reading and
writing of data transactions amongst distributed processes, without
losing control over the causality of sequences\cite{andras}.  Each
parallel shard or data handler $A_i$ can follow these process promises
locally for local spacetime consistency.

\section{The namespace and its pipeline semantics}

We can now turn to the the problem of integrating the totally ordered
writes of individual handler rings into a larger, partially ordered
set---sewing together and interleaving the results across a complete
spanning tree of the system. This is how a coordinatized namespace can
be curated, without consensus voting. The price for such a global
order is an increased uncertainty in the defined estimate of
`simultaneity', as parallel key writes span a larger and larger total
memory space. In other words, the wider the catchment area for data,
the coarser the temporal uncertainty of snapshots according to the
wall clock. This is an unavoidable and fully expected expression of
the usual Nyquist-Shannon sampling law, for any steady process (see
section \ref{uncert}).  As a side effect of deterministic correctness,
consensus is a once and for all operation that minimizes
communication. In a quorum solver, excess communication is required
to determine a value and communication is expensive (in terms of energy as well as latency).

\subsection{Hierarchical cooperative partial order}

Let's return to our client $C_a$, whose transaction job refers to key-value pairs
located across many $A_i$ from every part of the system. The client submits its 
transaction-related operations to individual handlers $A_i$ that own specific
$(k,v)$ pairs.

\beq
C_a &\imposition{+(k,v)}& A_i \in G(A)\label{transaction}\label{kv}\\
A_i &\promise{-k\in k(A_i)}& C_a\label{kvminus}.
\eeq

A scaled version of the local selection procedure has to keep the same
promises of the local solution: i.e. to only read or observe data from
the past cone globally, and to totally order parallel intervals (now
of increased duration). We therefore scale proper time using a
hierarchy of handlers, each with their own local time. A hierarchy is
an addressable structure across spacetime, which can be compared to
the addressable structure of the Internet's Routing or Domain Name
Services.  The more global the answers we need the higher up the
hierarchy we need to go to find root nodes with large scale indexing.
For actual data, we use the index to go straight to a local service handler $A_i$.

The scenario is still that shown in figure \ref{dbclock}. We divide up the
whole of spacetime into cells or rings of $A_i$ handlers, which all
perform an analogous function on a shard of mutually exclusive data.
Each agent works independently on its own data, but when it comes to
committing writes, the agents in a ring now take turns to interleave
their commits into the global records.  Each ring has a single parent,
which polls its children in turn to interleave finished writes. The
writes have been labelled with the proper time $\tau(A)$ of their own
storage process, but we have no idea how these are related to one
another for different $A_i$. Thus, the parent has its own version
counter (clock), which calibrates all the interleaved commits from the
children by committing its own record of the association. It collects
all keys written in a given handler time slot into a serial timeline
of its own, labelled by parent time (see figure \ref{interleave}). Each write happens locally
and immediately, but a globally accessible record of it may take some
time to propagate out to the whole of a space to form a complete
snapshot\footnote{The contentious aspects of availability have to do with the timescales
involved in processing. All data are subject to eventual consistency
for different observers, as they spread out into the future cone.}.

\subsection{Queueing commits for publication}

We can now walk through the mechanics of the process of publishing
committed writes explicitly\footnote{Note that nomenclature differs
  here between databases and version control systems.  We use `commit'
  in the sense of a database here, not in the sense of a version
  control system, where the term `push' might be used.}.  The service
agents form a collaborative hierarchy formed from edge handlers
$A_i$, parent handlers $P_j$ (see figure \ref{dbclock}), grandparents
etc.  These are collected into groups or rings at each level, which
interleave records from handlers at a lower level by pulling them in
batches from the queues assembled continuously by the handlers. This
secures a decoupled pre-emptive schedule for updating temporal order
at the next scale. At the handler level, all agents continue to
resolve data read and written locally, without waiting; however, a
token $\tau$ is passed around each ring by the parent.  The parent
shifts its from subordinate to subordinate to ensure fair interleaving
of data in a round robin fashion.  If a client is unavailable, the
parent can skip it temporarily while it recovers. As long as the cycle
spins with the same orientation, the integrity of the interleaving
will be maintained.

\beq
\text{define:} & j := (i+1)\text{\;mod\;} N(G) &\\
A_i &\promise{+(\tau+1) | D(\tau)}& A_{j}\\
A_{j} &\promise{-\tau | \neg \tau}& A_i
\eeq
The latter is equivalent to promising to commit transactions if and only if
the group proper time lies within the agent's appointed time slot.
\beq
A_i &\promise{+D(w_k)\; |\; \left( \tau(G) \;\in\; (i+\tau(G)\text{\;mod\;} N_G)\right)} &Q(A_i),
\eeq
Committed records are added to a log or queue $Q(A)$ by the handler, denoted
$Q(A_i)$, which is drained in batches by the parent whenever a child
has the coordination token. This pattern is replicated up the hierarchy.
At the top of the hierarchy, a root service node promises the global
time index, somewhat analogous to a root DNS record, or master routing table.
\beq
R &\promise{+(k,t(R),t(A_i))}& *.
\eeq
A purely voluntary scheme of cooperation is not pre-emptive, and can
easily be interrupted or sabotaged by a failed or floundering node.  A
pre-emptive approach can be secured by using each parent as the
arbiter of token ownership: since the purpose of the token is to
secure the exclusive attention of the parent: listening to agents one
at a time by {\em pulling} (uploading) a pre-packaged queue $Q(A)$ of
completed commits for the timeslot.

\beq P \promise{-D(w_k)\; |\; \left( \tau(G) \;\in\;
    (i+\tau(G)\text{\;mod\;} N_G)\right)} Q(A_i) \eeq The conditional
is now evaluated by $P$ not $A_i$.  No push-based (imposition) scheme
can be reliable due to the FLP result.  This could potentially add an
unwanted communication delay to hand over. However, this only affects
the slot for uploading (and thus publishing) commitments to $P$, not
their positioning within the schedule for updating the past cone,
which continues to be updated in parallel at best possible speed.
Thus, the serialization is only in the publication schedule, not in
actual data processing.  We can therefore examine the failure modes of
a single handler as a decoupled issue.

In general, secure behaviour by every $A_i$ has to rely on the fault
semantics of process handling by $A_i$, which is beyond the scope of
this discussion. We assume that $A_i$ does not break any promises to
the client or parent during its timeslot, in case of failure.  One way
to ensure this is to make updates based on synchronous reliable
transport\cite{paulentangle}.

The final resolution mechanism is roughly analogous to that used by
the DNS service for IP addresses, but unlike DNS the parents don't
reveal their clock versions downwards to synchronize $A_i$.  If the
$P$ offered their time as a service, this would add unmeasurable
delays, since $A$ doesn't need the value of P's clock as long as
service integrity is maintained (it would be similar to using NTP to
try to synchronize the clocks of the parallel clients, and with the
same flaws). Instead, the parent keeps the index association between
$k,v_n,t(a)$ and $t(P)$ as an aggregate log. This adds a scaling
burden, since the parent keeps records from all children and thus
needs to work faster to compress them into the same interval.  The
result is that the resolution of the past cone is eventually reduced
by greater spatial coverage. The interleaving is not linear, so we might not
notice for sparse probabilistic writes, whereas we might notice for
bursts of consistent updates. In any event, the timeline will be
ordered properly but the certainty of being able to see the latest
updates will become worse as traffic increases.

Once ordered, it's a simple matter to ensure a consistent published
view of shared data, by curating monotonically versioned streams,
labelled by data's origin or source handler $A_i(k)$.

\subsection{Key handler $A_i(k)$ semantics}

The edge handler lives at the edge of the network. Data need not be
transported from this edge except for long distance replication.  We
can now combine all the foregoing pieces into an algorithm at $A_i$:
\begin{enumerate}

\item An agent $A_i$ writes uncommitted records without delay, using the current local clock counter to record
the moment of writing. It's possible for several records to end up with the same timestamp. The channels' commit
policies promise to resolve such collisions and signal their acceptance of rejection of certain transactions.
\beq
C_a &\imposition{+(k,c)}& A_i\\
A_i &\promise{-(k,v)}& C_a\\
A_i &\promise{+w(k,v_\text{priv})| (k,v)}& C_a\\
A_i &\promise{+D(k,v,c) | \text{policy}(k,c), (k,v)}& P
\eeq
The parent node accepts the schedule timeline, where committed writes $(k,v,c,t_G)$ represents a version change
for the key $k$ in channel $c$.
The parent node proceeds to make the committed version available for expanding searches:
\beq
P \promise{+D(k,v,c)} *.
\eeq
These updates spread like ripples around the system, gradually becoming visible to all.

\item The index at $P$ associates a unique authoritative home $A_i$
  with a unique key for a data model, as well as a list of replicas by
  region.

\item A separate index per channel associates the graph of temporal order with key version updates, thus allowing
clients to walk through the data in curated temporal order.

\item If we want to be able to promise and recover temporal order over
  all locations, then we need a process such that the future cone is
  unique.
The parent node for $A_i$ maintains an index by group time $t_G$ and by key $k$ that will be published
upwards for global view, so that searches point to the authoritative handler for each key.
The parents thus store the authoritative order of records within their group.

\item As we go further up the hierarchy, the chance that records will need to order relative writes
on different $P$ regions grows progressively more unlikely. So the amount of sequential intent
should fall off appropriately.

It is up to a transaction handler to go as far up the hierarchy as necessary when writing correlated
sequences.

The parents can also checkpoint the $A_i$ clocks to its own to speed up searches of older data.

\item Once committed and published, independent replication processes can be used to speed up access
and maintain probable availability across large areas. Pull based replication (subscription allows greater
reliability and scale predictability than push based replica methods\cite{remi3}.

\end{enumerate}

\subsection{Parental Pre-emptive Bulk Indexing}

With a queue buffer we decouple publishing of committed records from
the actual act of commission, using voluntary cooperation as the link.
This offers both security and efficiency for the predictable
schedule\cite{burgessrpc,treatise2}.

\begin{figure*}[ht]
\begin{center}
\includegraphics[width=16cm]{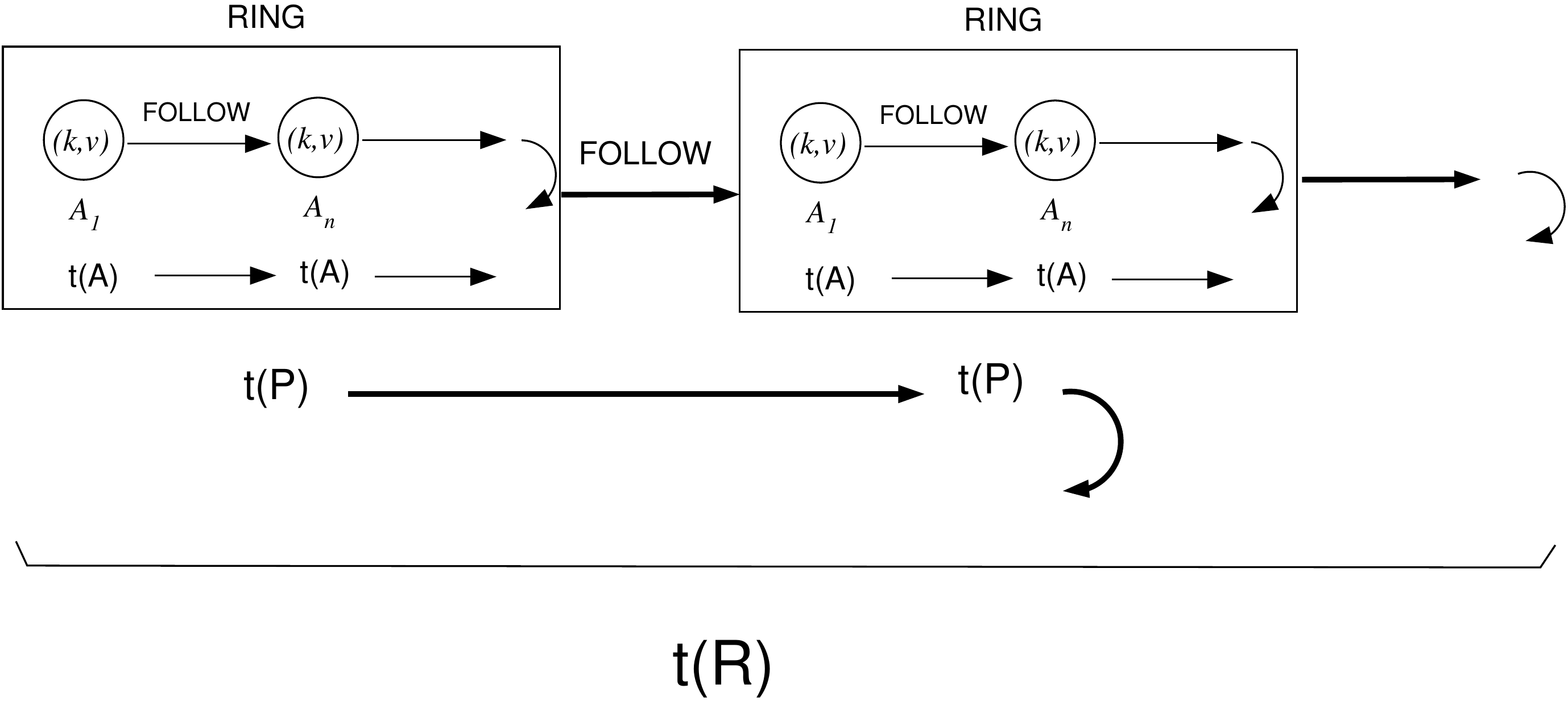}
\caption{\small The flow of aggregate batch updates through the hierarchies.
Updates from each cycle at scale $S^n$ are associated as a single batch
index with the timestamp of the parent scale $S^(n+1)$. This can scale
as far as the parent level can cope with the updates from the
children (which may involve sharding of indexing too). The size
of association records $\sigma$ is small compared to the original data.\label{dbclock2}}
\end{center}
\end{figure*}

The batch update schedule is as follows (see figure \ref{dbclock2}):

\begin{itemize}
\item At any time, agents write data $k.v_n$ to private workspace, where
$n=t(A)$. 
\item Each record in the committed chain of versions can be joined in
  a chain to its predecessor $(k,v_n) \promise{+\text{next}} (k,v_{n+1})$ to
  make search efficient (see section \ref{SST}). A graph representation

\item After collision avoidance, an agent $A_i$ adds a committed write
  to the queue $Q(A)$ as a record $\sigma_A = (k,t(A),c,i)$, which is
  complete pointer to $v_n$. Note that the record contains the agent
  for the key in short form $i$, since the sequence will not proceed
  without gaps for the timeslots.
\item As $P$ receives a batch $\Sigma_A$ of completed records 
with pattern $\sigma_A$, for the interval, it associates this batch with its
own timestamp: $\sigma_P = (t(P),\Sigma_A)$ and appends that to its data, and adds it to
its pipeline queue $Q(P)$ to be received by $R$.
\item As $P$ receives a batch $\Sigma_P$ of completed records with pattern $\sigma_P$,
it writes the association $\sigma_R = (t(R),\Sigma_P)$, and so on.
\end{itemize}

\subsection{Retrieving `latest' data from the hierarchy}

Unless specific version numbers are referenced in a search, we propose
the default search policy to be that: 
\begin{enumerate}
\item Any transaction refers to one
and only one version of a given key's value chain for its entire
execution, namely the last value written prior to the start of the
transaction. 

\item If a transaction alters values during the course of its execution, it
writes these changes to private versions of the that shadow the
initial value, to be committed at the end. Intermediate values are not
seen by other processes.
\end{enumerate}
The need to walk through every key, and thus reconstruct the entire past cone of a certain transaction
may be necessary. In order to find members of the causal past
from a given transaction, a transaction consults the root node
(perhaps via a directory service), which assesses the current starting $t(R)$
of the transaction, and uses this to parse the tree of associations.

Index services, for known key values, can enumerate the keys, find
their key handling shards $A(k)$, and pick the latest version
corresponding to $v_n \le t(R)$. Resolving this inequality quickly
will depend on the data indexing on the parents. Starting from the top
down is efficient since the ordered records of the parent hierarchy
associate all key changes with a single valued time, starting from $t(R)$.

All this depends on how often the version data are changing. Such a
confluence of searches on the index data can be shared by a farm of
search agents, since the data are read only during searches.

More commonly, transactions will want to know the latest consistent cone.
we need to parse the associations. This has to be quick.

Determining the precise order of any two versions belonging to different $A_i$,
we need to compare their timeslots in the relevant index:
\beq
v > v' \implies (v_R > v_R'), ~ (v_P > v_P'),~ (v_A > v_A').
\eeq

As far as lookups are concerned, the virtue of a hierarchy is that it
naturally load balances the lookup information as long as we know which node
to query in order to resolve times. However, we must also acknowledge that
treelike hierarchies are brittle structures and failures have to be offset
by redundant replication on all levels. Reliability on a massive scale is not cheap.

\begin{remark}[Synchronizing clocks with NTP?]
The difference between this scheme and simply deploying NTP on each
host is that we ask the $A_i$ to take turns in a monotonic round robin
cycle. This allows them to predict positions within a fairly
interleaved schedule. A small number of slots may be allocated between
the interleavings to ensure that a sudden burst would not block Thus,
this is a form of pre-emptive scheduling. No single host should be
allowed to block others, else the system becomes vulnerable to denial
of service attacks.
\end{remark}

\subsection{$n$-torus clock}

In the main presentation, we have used the ring structure for
interleaving parallel write processes. This is a key scaling process,
which benefits from a cyclical structure. Another process that can
benefit from a cyclic structure is data consensus and replication, which is often
handled by Paxos or Raft in contemporary thinking.  Rather than these
complex protocols, a simple approach would be to transparently use
reliable transport\cite{paulentangle} to pass copies of write
operations around a ring of replicas.  Once a data update returns, the
replication will be consistent and the ring acts as another
observability brake (like transparent copy on write). If we add this,
then we now have two rings, or a 2-torus:
\begin{enumerate}
\item Shard ring (concurrent interleaving).
\item Consensus ring (copy on write).
\end{enumerate}
With both of these dimensions, after a complete cycle in both directions, we
know that data writes are replicated and ordered for publication.
Indeed, other constrained processes, like multiple schema collections,
models, and tabular types, can be decoupled in a similar way, leading
to an $n$-toroidal clock. The motivation for this is to avoid
unnecessary waiting. It can be shown that the two dimensions above are
sufficient however.

The precise behaviour of a data pipeline must be attuned to the
particular application use-case in question. In current thinking,
databases are generic contraptions tuned for one and only one mode of
usage. This is one reason why we need so many different kinds of
database. However, if users `stated their intentions' as policy up
front, then the backend system could accommodate and interleave
multiple needs. This kind of thinking is growing in popularity.

\section{Scalability of the clock}

There are many open questions in describing the proposed service model,
but we can make some remarks about the scaling rates.

\subsection{Granularity of clock layers}

Our data pipeline `clock' is driven from the bottom up, client
requests determine the amount of data flowing up to the root nodes.
The size of the data propagating up the pipeline is a concern for the
stability of the service. The application context will determine the
stress on the system for `mostly write' or `mostly read' regimes.

As associative timestamps move up the pipeline, no actual data need to
be moved, so each association is a small amount of data. However, each
layer packs in data from parallel sources, so the total amount may
still be large.  Moreover, the timespan (as perceived in client time
$\tau(C)\sim t(A)$) of each ordered batch or uniquely identifiable snapshot
grows by an order of magnitude for each layer of aggregation.
The `thickness' of the coarse grains varies like
\beq
\frac{dt(R)}{dt(A)} \simeq N_A \times N_P \times \ldots,
\eeq
where $N_A$ is the number of hosts per $A$-group, and so on.
We can't know or calculate the exact times without a complete
specification, but if we assume that each aggregation collects
an order of magnitude $|G|$ then
\beq
\Delta t(R) \simeq \Delta t(R) \times |G| \times |G|\ldots,
\eeq
or for $\ell$ layers,
\beq
\frac{dt(R)}{dt(A)} \simeq \prod_i^\ell |G_i|
\eeq
So, if a client process takes milliseconds, then a three layer hierarchy
with 10 shards over three regions, would mean a time resolution of around 30 milliseconds.
This is a rough estimate, but it indicates that the serialization processes are not free.

\subsection{Spacetime sampling resolution}\label{uncert}

A final point about the coarse graining mentioned in the previous section.
We can interpret this as the Shannon-Nyquist theorem's fundamental limit on the
observability of periodic spacetime processes\cite{shannon1,cover1}.  For steady state
processes, we can decompose update cycles into Fourier series and
compute a fundamental uncertainty relation. This corresponds to the
well known Heisenberg uncertainty in quantum
mechanics\cite{heisenshannon}: \beq \Delta \omega \Delta t \ge S \eeq

\begin{lemma}
We can define the past cone for a record with varying levels of precision
in a scale-dependent way, because the accuracy in time is proportional to
confinement in space.
\end{lemma}
We define the maximal past cone for a transaction to be that generated by records in $t(R)-\Delta t(R)$ so we
need and indexing function that takes a current time for a service handler and returns
the head of its past cone:
\beq
f(k,t(A)) \promise{\text{returns}} (k,t(R)),
\eeq
since the sequential log starting from $t(R)$ points to all subordinate versions across its
spatial catchment area, by virtue of the single aggregation pipeline's process gradient.
\begin{lemma}
All this amounts to a coarse grained foliation into spacelike hypersurfaces, where the thickness
of the surfaces is determined by the cycle period at each level.
\end{lemma}

$R$ points to the head of the past cone for each key's write history,
with associated inputs from $t(A), t(P), ...$. In practice, a complete
spanning tree of associations is only needed when reconstructing a
search that involves multi-key temporal order. We can find single
values quickly from their handlers. In other words, when involving
multiple keys with times that were correlated over spatial
separations, the spacetime trajectory of the past-cone updates has to
use a shared proper time as the search parameter. By contrast, the temporal order
for a single key is always determined by A alone.

When searching for the past cone, each handler may thus have a chain of
versioned indices over its children to traverse. In order to find an index
that covers the complete observable cone of the process, starting
from $A_i$ a broken thus follows the chain until it reaches a node in
$S^{n}$, whose clock presides over all the subordinates.  

By the downstream principle, a receiver of data is always the arbiter
of its interpretation: i.e. its value, its time stamp, etc. So, if
some $P$ sends to $A$, then then $A$ decides its current value.  If
$A$ sends a value to $P$ then $P$ decides its current value.  The
assumption in using higher levels to adjudicate time for lower levels
is that the delays incurred by association are unavoidable.  This may
be hard to accept for engineers who still believe that latency is a
bug rather than a feature of communication; however, one cannot make
progress without accepting this as an unavoidable fact of
communication. The best any system can promise is that the publishing
of committed results moves from bottom to top monotonically, and thus
the fully shared interpretation of absolute past (however delayed) is
only published as a single version for all of space (at the same
logical moment) by the top node.

\subsection{Self-governing rate-limiting for thrashing protection}

As with the self-governing, rate-limiting architecture in CFEngine\cite{lisa97113},
each agent's promises ensure that it is immune to client traffic bursts.
We make handler acceptance conditional on the work queue, i.e.
handlers accept transactions provided the write commit queue is not too long:
\beq
A_i \promise{-T \; \Big|\; |Q(A_i)| \;<\; L } C_a,
\eeq
where $|Q(A_i|$ denotes the length of the queue, and $L$ is some configurable
policy for maximum length.

The interleaving of data by the parent is essentially a packing
problem. In average operation, writes may be relatively sparse and
packing is conservative. However, busy data pipelines may integrate
large amounts of regular data from a known list of sources leading to
a dense and continuous stream of writes.
The most challenging integration problem will be when every thread is
writing new data in parallel to different keys, so that each parent
update involved in the interleaving is fully packed. This could
overload the input queue of the parents. However, one could brake the
process naturally by waiting for the parent records to be written on
commit before ending a session. Rate limiting of connections may be
necessary in any service to avoid queue divergence.

This packing process places a burden particularly on the root nodes of
the hierarchy. The flow limitations implicit in this packing make the
root node (including its replica set) the ultimate choke point for data
flows. If they can't keep up with write versioning, the global index
will fall behind.  By ensuring the emptying of the queue by pulling,
and the refusal of connections when the queue is full, this limitation
propagates back down as an autonomic throttle for self-protection.

\begin{remark}
If key changes are in different shard handlers $A_{j\not=i}$ then
  they can still occur in the same public time interval, i.e. within
  the same coarse time tick, but if they are repeated keys, they must
  be serialized, so we need a consistent way of serializing locally,
  and aggregating in parallel---quickly, and with no limit on spatial
  separation. These changes can be labelled in the same version intervals,
but we have no way of knowing when clients will see these update, because
observation depends on both the promises of publishing (+) 
by the service and subscribing (-)  by the client.
\end{remark}

The parent only has to receive data that are finally committed to different channels.
This may be bursty. These updates don't need to fall in the trap of push updating.
They can be queued on the $A_i$ themselves and be pulled at the parent's convenience
with a fair interleaving policy\cite{koalja}, like a managed data pipeline, thus avoiding
the perils of push imposition semantics. Since publication of committed transactions
is performed by the parents, there is no danger that data will become unsynchronized.
Moreover, as long as the number of channels is limited, there will be a natural rate limiting
effect from handling contention

For example, an agent could write to a different channel at a future
time, but is not allowed to win a `latest' race by teleporting into
the future. These are the semantics of the default channel.

\section{Discussion etc}

In this paper we've written down a deterministic interleaving process
for clocking proper process time over a spanning network of data
handlers\cite{andras}, in which the propagation of key value versions
is automated, following the principles well known in Internet address
scaling. The role of our hierarchical `clock' is to assign unique but
distributed coordinates to what amount to CRDTs, thus decoupling the
processes in a data lifecycle: capture, replication, committing,
publishing, and searching, and ensuring that every search will always
see the consistent causality cone intended by the writer.

Much of the reason for trouble in handling the scaling of time order
lies in the almost universal adoption of push based semantics, which
are unreliable and expensive to adapt, and the belief that such operations
are instantaneous (impositions in Promise Theory language) as
described by (\ref{kv}). In a push scheme, data arrive in an
uncontrolled manner at a service point, where an escalation of
resource ensues to try to cope: increased queue lengths, load sharing
dispatchers, intermediate agents, etc. The rational alternative to
this is the Publish-Subscribe or Pull based methods used by Content
Delivery Networks, where directory services assign queue handlers, and
clients help-themselves in a self-service lookup.

The alternative algorithmic implementations of a global proper time,
described earlier, use transported tabular memory to count ticks with
`tensor clocks', or tables that accumulate path history as the process
propagates from agent to agent by virtual motion\cite{virtualmotion}.
Scalar, vector, and even matrix versions of such clocks have been used
for labelling proper time along paths through
networks\cite{vectorclocks,plausibleclock,intervalclock,fridge}. These
provide one solution to partial orderings with single threads, but
they are expensive in terms of communication, which is the major cost
in scaling, because they grow monotonically in size, and don't resolve
all the ambiguities of distributed counting. In a sense, our
alternative uses a distributed counter like a stigmeric process.

We anticipate this coordinate namespace to be especially important in
highly distributed applications, such as edge computing, including the
development of the infamous services as part of 5G and 6G wireless
networks. Maintaining speed and locality of service, but with far
reaching access, is the optimal way to scale data services. Today,
long distance data flows are a major cause of IT's carbon footprint,
which is unsustainable for the expected growth. In that sense, what we
describe here is a universal infrastructure for data services, which
scales both predictably and as reliably and efficiently as can be with
current technology. The approximate intuitions of the so-called CAP
Theorem are then surpassed by a mixture of deterministic correctness
and redundant best-effort availability, thanks to our abandonment of
strong coupling quorum semantics.

One should never underestimate the cost of maintaining structural
information on a large scale. There is no free lunch, as they say. A
partition is a partition. Redundancy is costly, and intrinsic latency
is unavoidable. In this scheme, garbage collection of temporary writes
(CRDTs) will be an important part of keeping a data field operational
for long durations. This is another under-addressed problem, though in
our cyclic design, this is a simple matter to automate through
scale-dependent policy.  Our aim, with this composite approach, is to
minimize communication and unnecessary transport of data, which both
have a high energy cost.  The ability to replicate state with `Just In
Time' streaming of just the necessary dependencies is a realistic
alternative to extensive caching, and this could be optimized with
machine learning enabled smart caching. Many details remain that we'll return to 
in future work.

Parts of the discussion in this paper are embodied by the patents
US K1093PCTUS and EU EP3794458A1.

\appendix

\begin{figure*}[ht]
\begin{center}
\begin{tabular}{|c||l|}
\hline
\sc Symbol & \sc Description / Meaning\\
\hline
$G_A=\{A_0,A_1,\ldots A_{N-1}\}$ & The set of handler agents forming group $G_A$.\nonumber\\
$N(G)$ & Number of agents $A_i$ in group $G$\nonumber\\
$k$ & An intended primary key name for a data record (always invariant)\nonumber\\
$A(k)$ & The invariant handler agent for key $k$.\nonumber\\
$v=v(k)=\{v_1,v_2,\ldots\}$ & A data value for key $k$\nonumber\\
$n(v_n)=n$ & The most recent (head) version of $v(k)$.\nonumber\\
$c$ & A data channel. The default channel is `latest'.\nonumber\\
$D(w_1,w_2,\ldots,w_k)$& A data registration or `commit' operation finalizing a number of write operations $w_k$.\nonumber\\
$w_k=w(k,v,c)$ & A write of value $v$ to key $k$ into the channel $c$.\nonumber\\
$r_k=r(k,v,c)$ & A read of value $v$ to key $k$ from the channel $c$.\nonumber\\
$\tau(G)$ & The current value of the exterior, shared proper time for group or agent $G$.\nonumber\\
$t(A)$ & The current value of the private proper time at agent $A$, \nonumber\\
     & which is related to $t(A)=(i + \tau(G) \mod N)$.\nonumber\\
$t(P)$ & The current value of the private proper time at agent $P$, \nonumber\\
$t(R)$ & The current value of the private proper time at agent $R$. \nonumber\\
$\sigma_i$ & A clock association for a key update, a tuple $(k,t(A),t(p),v_n)$\nonumber\\
$\Sigma_A$ & A commit set of associations $\sigma_i$ passed to a parent from $A_i$\nonumber\\
\hline
\end{tabular}
\caption{\small Summary of notation used in the paper. Notice that a single commit operation $D(w)$ can only
contain each key once only.\label{summary}}
\end{center}
\end{figure*}

\subsection{Promises}

\begin{figure}[ht]
\begin{center}
\includegraphics[width=7.5cm]{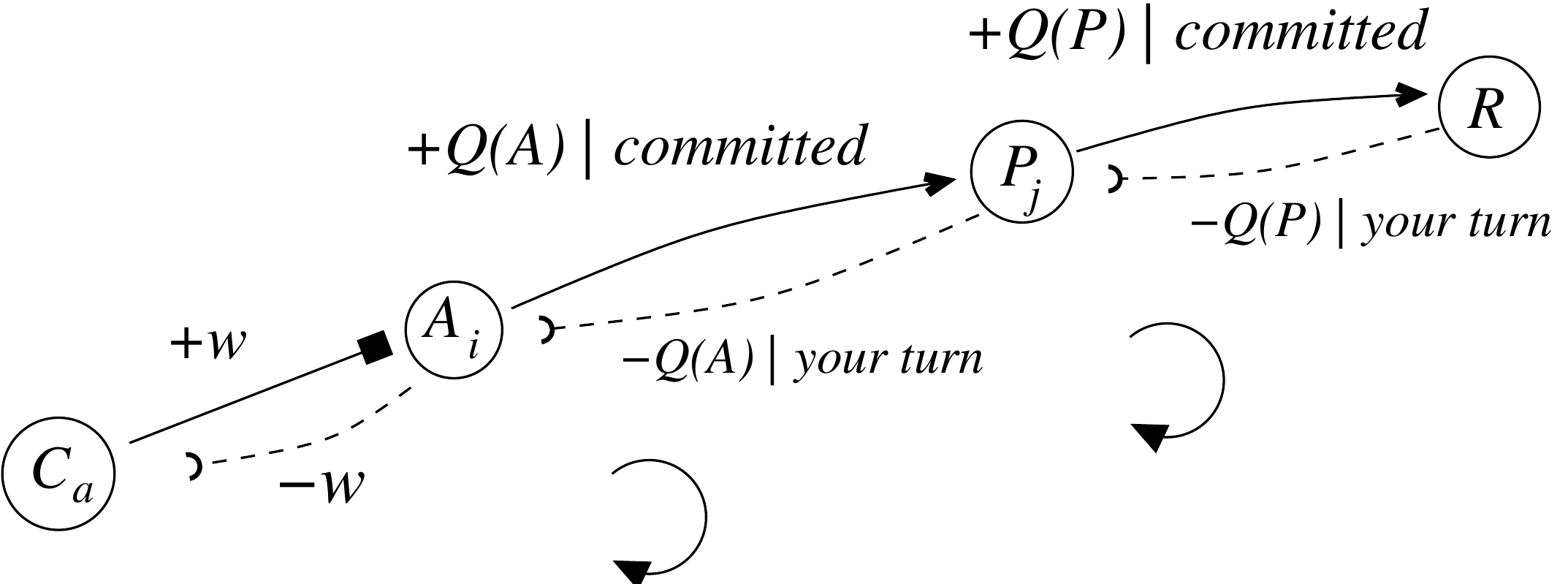}
\caption{\small The schematic pattern of service-offer promises within the hierarchy of agents.
Clients push or impose writes onto a handler. The handler performs writes and schedules
the slots for committing the results within a round robin collection scheme. The parents pre-emptively
empty the handler queues in round robin fashion, and associate these with its own clock to calibrate them.
The parents calibration records are scheduled for commitment to the next level, and the parent of the parent
empties this queue when its timeslot comes around at the next level, and so on.\label{promise}}
\end{center}
\end{figure}

We summarize some notations and terminology for convenience in figure \ref{summary}.
We use the notation of \cite{promisebook} in which an agent $A$ makes a promise
to another agent $A'$ with body $b$, which describes the nature and magnitude of the promise:
\beq
A \promise{+b} A',
\eeq
where a $+$ sign denotes an offer. An imposition is an attempt by $A$ to opportunistically
induce a response in $A'$, without prior warning, and is written with a `fist' arrow:
\beq
A \imposition{+b} A',
\eeq
If the subject of the promise or imposition is accepted by $A'$, a corresponding $-$ promise
is given, denoting a causal binding from $A$ to $A'$:
\beq
A' \promise{-b'} A.
\eeq
The extent of the coupling is $b \intersection b'$. If a promise of $b$ is conditional
on the receipt of promised conditions $c_1,c_2,\ldots$ then we write
\beq
A \promise{+b | c_1,c_2\ldots} A',
\eeq
The arrow notation is convenient to draw in figures, such as figure \ref{promise}.

Every shared service wants to utilize its resources as efficiently as possible by multiplexing
shared flow bottlenecks in an optimal way. Thus key-handlers promise fair policy-based multiplexing
for access to keys.

\subsection{Definitions and scales}

In keeping with other literature on databases, consistency, and snapshot isolation,
we define the following terms:
\begin{definition}[Data operation]
An operation is either a `read' or a `write' of a single key-value pair. 
\end{definition}
\begin{definition}[Commit]
Data writes are local and private (like local variables in software). Changes are said to be
committed when they change status to `published' in a shared stream. The commit operation
is authoritative and assigns the approved successor in a key-value version stream. Commit
operations are mutually exclusive in space $k$ and time $\tau$.
\end{definition}
\begin{definition}[Channel or data history]
  A history is a sequence of data write operations, also called a trajectory
  in semantic spacetime parlance. Successive values in a history form a channel.
Several alternative channels may be maintained with different identities and semantics.
\end{definition}
The concept of a transaction is often confusing to 
those who don't work in database jargon. A transaction is not an atomic write, but
something closer to a closed function in programming.
\begin{definition}[Transaction]
  A protected `quasi-atomic' unit of work, or a single command issued
  to a data service, which may be viewed a sequence of data operations
  and ends with a `commit' operation.  A transaction is sometimes
  considered to be reversible, though we shall not assume this.
\end{definition}
Note we say quasi-atomic because transactions are not indivisible, they are only wrapped
as black box changes, whose interior machinations are unobservable.

\subsection{Timescales}

Time plays a key role throughout any communication system. We deal
with not only a variety of processes but also a variety of typical
scales for interaction.
Clocks are used almost everywhere for computation, but the common
notion of time as what we see on the wall clock (or UTC in the digital
era), is not the clock we need. A clock is any reference process that
we use measure and calibrate change.  A clock has three major
functions:
\begin{itemize}
\item To resolve the partial order of events by counting faster than 
the phenomenon of interest, according to the Nyquist law.
\item To define the cone of absolute past, and 
\item To be able to reconstruct a snapshot of the current spatial hypersection, which is
what we understand as a `snapshot' of the data.
\end{itemize}

Proper time is the time counted by an agent that executes the steps of
a local process with its interior resources\cite{treatise2}. When processes
are distributed, the counting of proper time moves from agent to
agent, along the trajectory of the process (as with vector clocks). We
may further define this to be the `intended' order of events since it
is co-moving with the origin process and represents the order in which
data were sampled and submitted for archiving---straight from the
process's mouth, so to speak\cite{steinke2002unified}.
Is the time defined by the recipient of the data or the imposer of the data, i.e.
compare expressions (\ref{kv}) and (\ref{kvminus}).
This was referred to as GDO and GWO in the language of \cite{steinke2002unified}

We define $t(A_i)$ to be the value of an atomic counter at an agent $A_i$, which increments
for every change operation, and thus represents its proper time.
If lost, the counter could be recovered by a record scan (analogous to fsck in Unix).

Orders of magnitude in terms of UTC time are helpful to gauge the effect of
a distributed system on the human world. For example, the timescales for common operations:
\begin{enumerate}
\item A direct indexed read, taking milliseconds $C \imposition{+r} A$.
\item A direct indexed write, taking milliseconds  $C \imposition{+w} A$.
\item A search and computation leading to reading multiple values, taking hours, involving many agents.
\item A search and computation leading to writing multiple values, taking hours, involving many agents.
\end{enumerate}
The range of timescales over which searches and computations preside
represents a challenge for giving meaning to multiple process channels
in the face of parallelism.  While reads and writes take only a short time, 
complete transactions that rely on a consistent view of data make persist
for many hours.

In general writing data is more time consuming than reading data,
since reads can be parallelized, while writes to a common location can
only be performed serially as a FIFO queue.  Thus, some processes are
optimized for mostly-writing data, some for mostly reading past data,
some always want the latest value (even during a storm of updates),
and various admixtures of these extremes.  

When many short data writes overlap with a long sequence of reads that
compute a result, the changes to the read set could potentially skew
the result.  Whether this is right or wrong depends entirely on the
{\em intent} of the data operation.

In our system for interleaving data with `round robin' scheduling,
there are cycles used to coordinate relative order by shared timekeeping. The time
allocated to each computer in a cycle before handing over to the next
is (in principle) a configurable policy choice. We want this cycling
to be at a quick steady rate for interleaving busy transaction queues.
Unlike a wall clock, a version system doesn't need to count artificial
changes to measure an independent time in between changes: it only
needs to increase the shared time counter when a change arrives.
However, no harm is done by counting at a regular rate, since records will
typically record UTC time for convenience if they need to.

\subsection{Spatial scales}

Space is a representation of memory in a computation. Each addressable location
involves physical and logical extent. We are interested both in geo-spatial
extent (which informs us about physical latency) as well as the space of
logical key values, which is relevant algorithmically.

In semantic spacetime, locations are agents that can keep promises,
i.e. exhibit functional behaviour. We can think of them as software
programs or self-contained processes.  Logical or semantic scales form
a hierarchy, as levels denoted by $S^{n}$ in semantic
spacetime\cite{spacetime2}. A collaboration of agents at scale $S^{n}$
may be thought of as a single agent at scale $S^{n+1}$.

At the edge of the network (level $n=0$), the collection of all
clients processes may span the entire globe or beyond. However, they
interface with a much smaller number of service points which are
scattered around the catchment areas.

Some databases use front end `controller nodes' to mediate a connection with 
specialized shard handlers. Equally, client libraries could mediate the connection
directly from the client side. Thus, in this note, we do not discuss the role of the
client or service brokers, as we are concerned mainly with the behaviour of shard handlers.

A local group of handlers $A_i$ is written $G(A)$ and is assumed to be co-located on the scale of a single
datacentre, while parent nodes may exist at a datacentre level, grouped across a city at
level $G(P)$, to a region level, a country level, and so on.  There's no limit,
in principle, to the scale to which a system might expand---though
eventually the burden of tracking a single process must grow beyond
acceptable limits.

Naming often gets in the way of more general progress: some may refer
to `scaling up' versus `scaling out', horizontal versus vertical
scaling, and so on.  These terms serve to distinguish meanings in a
local context, but general principles are often elusive.
The term `top down' means from a global view of time, and it starts
from the `root' node for a span of the system.  Conversely, `bottom
up' means a local view of time, which is begins at the edge or leaf
nodes of the system span.  What one would refer to as scope in
software engineering, for process encapsulation of operations and
transactions, is sometimes called an isolation level is database
parlance.

\subsection{Semantic spacetime}\label{SST}

Semantic spacetime is constructs processes from four basic spacetime relations:
i) FOLLOWS (for temporal order), ii) CONTAINS (for spatial aggregation), iii) NEARness for
location or semantic equivalence, and iv) EXPRESSES for scalar value expression.
The structure of a namespace coordinates can be summarized for our causality cone with
the following promises:

\beq
A_{i+1} &\promise{+\text{\sc FOLLOWS}}& A_i ~~\text{(rings)}\nonumber\\
v_{n+1} &\promise{+\text{\sc FOLLOWS}}& v_n ~~\text{(versions)}\nonumber\\
\eeq
\beq
A_i(k) &\promise{+\text{\sc CONTAINS}}& k ~~\text{(key-value)}\nonumber\\
(k,v) &\promise{+\text{\sc EXPRESSES}}& \{ v_n\} ~~\text{(version strings)}\nonumber\\
A_{i}^\text{cache} &\promise{+\text{\sc NEAR}}& A_i ~~\text{(replicas)}\nonumber\\
\Sigma_{A_i} &\promise{+\text{\sc CONTAINS}}& \sigma_i(k,v_n,t(P)) ~~\text{(simultaneous)}\nonumber\\
\sigma_i \in \Sigma_A &\promise{+\text{\sc NEAR}}& \sigma_i \in \Sigma ~~\text{(simultaneous)}\nonumber\\
T_1(w(k)) &\promise{+\text{\sc NEAR}}& T(w(k))~~\text{(overlapping $T_1$)}\nonumber\\
\eeq

\bibliographystyle{unsrt}
\bibliography{spacetime,spacetime2}

\end{document}